\documentclass{aa}  
\usepackage{graphicx}
\usepackage{txfonts}
\usepackage[colorlinks=true]{hyperref}
\hypersetup{linkcolor=blue, citecolor=blue}
\usepackage[draft]{todonotes}

\begin{document} 

   \title{CO enhancement by magnetohydrodynamic waves}
   \subtitle{Striations in the Polaris Flare}
    \titlerunning{CO enhancement by magnetohydrodynamic waves}
        
   \author{R. Skalidis\inst{1}\thanks{skalidis@caltech.edu}, K. Gkimisi\inst{2} \fnmsep \inst{3} , K. Tassis\inst{2} \fnmsep \inst{3}, G. V. Panopoulou\inst{4}\fnmsep \inst{5}, V. Pelgrims\inst{2} \fnmsep \inst{3}, A. Tritsis\inst{6}, \and P. F. Goldsmith\inst{7} 
          }
          
     \institute{
    Owens Valley Radio Observatory, California Institute of Technology, MC 249-17, Pasadena, CA 91125, USA 
    \and
    Department of Physics \& ITCP, University of Crete, GR-70013, Heraklion, Greece
    \and
    Institute of Astrophysics, Foundation for Research and Technology-Hellas, Vasilika Vouton, GR-70013 Heraklion, Greece
    \and
    Department of Physics, University of Cyprus, 1 Panepistimiou Avenue, 2109 Aglantzia, Nicosia, Cyprus 
    \and 
    Department of Space, Earth \& Environment, Chalmers University of Technology, SE-412 93 Gothenburg, Sweden
    \and
    Institute of Physics, Laboratory of Astrophysics, Ecole Polytechnique F\'ed\'erale de Lausanne (EPFL), \\ Observatoire de Sauverny, 1290, Versoix, Switzerland 
    \and 
    Jet Propulsion Laboratory, California Institute of Technology, 4800 Oak Grove Drive, Pasadena, CA 91109-8099, USA
    }

   \date{Received; accepted}

   \newcommand{\rskal}[1] { {\color{purple} #1} }

  \abstract
   {The formation of molecular gas in interstellar clouds is a slow process, but can be enhanced by gas compression. Magnetohydrodynamic (MHD) waves can create compressed quasiperiodic linear structures, referred to as striations. Striations are observed at column densities where the atomic to molecular gas transition takes place.}
   {We explore the role of MHD waves in the CO chemistry in regions with striations within molecular clouds.}
   {We target a region with striations in the Polaris Flare cloud. We conduct a CO J=2-1 survey in order to probe the molecular gas properties. We use archival starlight polarization data and dust emission maps in order to probe the magnetic field properties and compare against the CO morphological and kinematic properties. We assess the interaction of compressible MHD wave modes with CO chemistry by comparing their characteristic timescales.}
   {The estimated magnetic field is 38 - 76 $\mu$G. In the CO integrated intensity map, we observe a dominant quasi-periodic intensity structure, which tends to be parallel to the magnetic field orientation and has a wavelength of one parsec approximately. The periodicity axis is $\sim 17 \degr$ off from the mean magnetic field orientation and is also observed in the dust intensity map. The contrast in the CO integrated intensity map is $\sim 2.4$ times larger than the contrast of the column density map, indicating that CO formation is enhanced locally. We suggest that a dominant slow magnetosonic mode with estimated period $2.1 - 3.4$ Myr, and propagation speed $0.30 - 0.45$ km~s$^{-1}$, is likely to have enhanced the formation of CO, hence created the observed periodic pattern. We also suggest that, within uncertainties, a fast magnetosonic mode with period 0.48 Myr and velocity $2.0$ km~s$^{-1}$ could have played some role in increasing the CO abundance.}
{Quasiperiodic CO structures observed in striation regions may be the imprint of MHD wave modes. The Alfv\'enic speed sets the dynamical timescales of the compressible MHD modes, and determines which wave modes are involved in the CO chemistry.}

    \authorrunning{Skalidis et al. 2022}
    \maketitle

\newcommand{\katia}[1]{\textcolor{red}{(katia: #1)}}
\newcommand{\LC}{$\Lambda_{\rm C^{+}}$}
\newcommand{\MS}{$\mathcal{M}_{s}$}
\newcommand{\MA}{$\mathcal{M}_{A}$}
\newcommand{\PlasmaBeta}{$\mathcal \beta$}
\newcommand{\VA}{V$_{\rm{A}}$}
\newcommand{\Bpos}{B$_{\rm{POS}}$}
\newcommand{\dtheta}{$\delta \theta$}
\newcommand{\PsiSV}{$\psi_{S/v}$}
\newcommand{\nHmol}{$n_{\rm{H_{2}}}$}
\newcommand{\NHmol}{$\rm N_{\rm{H_{2}}}$}
\newcommand{\NHtot}{$\rm N_{H}$}
\newcommand{\EBV}{E (B-V)}
\newcommand{\dthetaIntrinsic}{$\delta \theta_{0}$}
\newcommand{\NCO}{N$_{\rm{CO}}$}
\newcommand{\ICO}{I$_{\rm{CO}}$}
\newcommand{\TK}{$\rm T_{K}$}
\newcommand{\XCO}{$\rm X_{\rm CO}$}
\newcommand{\depth}{$l_{0}$}

\newcommand{\mum}{$\mu$m}
\newcommand{\muG}{$\rm{\mu G}$}
\newcommand{\kms}{km~s$^{-1}$}
\newcommand{\ColDens}{cm$^{-2}$}
\newcommand{\VolDens}{cm$^{-3}$}

\newcommand{\HI}{H\,{\sc i}}
\newcommand{\CII}{C$^{+}$}
\newcommand{\CI}{[C\,{\sc i}]}
\newcommand{\Hmolecular}{H$_{2}$}
\newcommand{\CO}{$^{12}$CO}
\newcommand{\COThirteen}{$^{13}$CO}

\newcommand{\tcool}{$\tau_{\rm cool}$}
\newcommand{\tslow}{$\tau_{\rm slow}$}
\newcommand{\tfast}{$\tau_{\rm fast}$}
\newcommand{\tco}{$\tau_{\rm CO}$}
\newcommand{\tmol}{$\tau_{\rm H_{2}}$}

%
\section{Introduction}

Molecular hydrogen is an important ingredient for the formation of stars. Molecular gas cannot survive in regions of low extinction because it is rapidly photodissociated by the interstellar (ISM) ultra-violet (UV) background of our Galaxy. Molecular gas forms in local dense regions within atomic clouds which are shielded from the UV radiation by the outer layers of the cloud \citep{Snow2006}. 

The formation of molecular gas is slow, compared to the estimated cloud lifetimes \citep{tassis_2004, mouschovias_2006, goldsmith_2005, goldsmith_2007, meidt_2015, murra_2011, Chevance_2022}, and scales with density (n) as $\sim 10^{3}~\rm {Myr} /n$ \citep{goldsmith_2005, Chevance_2022}. This means that an atomic cloud with initial density $\sim 10$ \VolDens, would become molecular in 100 Myr.

The timescale of the \HI\ $\rightarrow$ \Hmolecular\ transition can be shortened via compression of the gas \citep{glover_2012, levrier_2012, Valdivia_2016, gong_2017, gong_2018, gong_ostriker_2020, bialy_2017, seifried_2016, seifried_2017}. Compressible flows induce local density enhancements, where self--shielding of the gas becomes more efficient. As a result, the \Hmolecular\ formation is accelerated within compressed regions. Direct numerical simulations of compressible turbulence show that the \Hmolecular\ formation timescale can be shortened to a few Myrs \citep{glover_2007a, glover_2007b}. 
Compressible flows tend to form filamentary gas structures within simulated clouds \citep{pudritz_2013, checn_2014, xu_2019_filaments_striations, Federrath_2021}. These filaments are the laboratories where molecular gas forms at an enhanced rate \citep{walch_2015}. 

Filaments have been widely observed in ISM clouds \citep[e.g.][]{andre_2010, clark_2014}. Filaments exhibit a wide range of properties, spanning orders of magnitude in mass, column density and length \citep{hacar_review_2022}. A specific sub-class of filaments, termed striations, has been found at column densities $\log{\rm N_{H}} = 20 - 21$ \ColDens\ \citep{goldsmith_2008} in the outskirts of nearby molecular clouds. These column densities match those where the \HI\ $\rightarrow$ \Hmolecular\ transition occurs \citep{Gillmon2006, bellomi_2020}. This paper examines whether striations are causally related to the formation of molecular gas.

Striations are elongated structures located within the diffuse parts of molecular clouds \citep{goldsmith_2008}, and are characterized by two main observational properties: 1) quasi-periodic fluctuations in column density maps \citep{goldsmith_2008, palmeirim_2013_striations, malinen_2014_striations, malinen_2016_striations, cox_2016} with contrast up to $\sim 25 \%$ \citep{tritsis_2016}; 2) the major axis of striations is aligned with the plane-of-the-sky (POS) magnetic field orientation \citep{heyer_2008, cox_2016, chapman_2011_taurus, malinen_2016_striations}. 

The observed alignment between the magnetic field morphology and striations, suggests that magnetic fields are important for the formation of striations. \cite{tritsis_2016} proposed that striations are formed due to the propagation of compressible magnetohydrodynamic (MHD) waves (also \citealt{beattie_2020_striations}). Key predictions of the \citealp{tritsis_2016} model have been verified in observations \citep{tritsis_2018_sc, tritsis_2018, tritsis_2019}. Alternative models suggest that striations could be formed due to the Kelvin-Helmholtz instability \citep{heyer_2016}, due to corrugations in magnetized sheet-like structures \citep{chen_2017_striations}, or due to anisotropic turbulent phase mixing \citep{xu_2019_filaments_striations}.

If striations are formed by compressible MHD waves, then the latter are likely to play some role in the \HI\ $\rightarrow$ \Hmolecular\ transition. Magnetic fields affect the dynamical properties of striations and thus may also affect their chemical properties. Here, we test the hypothesis that the formation of striations and molecular gas formation are linked.

We targeted the Polaris Flare cloud, which is a diffuse cloud with prominent striations \citep{andre_2010, panopoulou_2015, panopoulou_2016}, at column densities typical of the \HI\ $\rightarrow$ \Hmolecular\ transition.
The plane of the sky (POS) magnetic field, as traced by stellar polarization, is parallel to the striations, and its strength was estimated to be 24 - 120 \muG\ \citep{panopoulou_2016}. 

We revisit the analysis of the magnetic field properties in the Polaris Flare striations. We present an improved analysis of the optical polarization survey data from \cite{panopoulou_2015}, using machine learning methods to increase the sample size of stellar polarization measurements. We use these new data to re-estimate the magnetic field strength. For the first time, we measure the kinematics of striations at high spatial resolution by mapping the CO (J=2-1) line and compare the magnetic field properties with the morphology and kinematics of the region. We compare the characteristic propagation timescales of MHD waves in the Polaris Flare with the CO formation timescale. We find that density waves (slow magnetosonic modes), propagating a few degrees off from the mean magnetic field orientation, could have enhanced the CO formation there, forming structures perpendicular to the magnetic field. Here, we adopt the following nomenclature: a) we refer to the target region as the "striations region", b) striations refer only to dust intensity structures; we distinguish the CO from the dust intensity structures, except otherwise mentioned.

This paper is organized as follows. In Sect.~\ref{sec:co_observations}, and~\ref{sec:polarization_data} we present the CO and polarization data. In Sect.~\ref{sec:co_striations} we show the CO kinematics and morphological analysis. In Sect.~\ref{sec:bpos_properties} we present the POS magnetic field properties (morphology, and strength), and discuss how it correlates with the CO properties. In Sect.~\ref{sec:striations_CO_chemistry} we discuss the potential impact of the linear MHD modes in the CO chemistry of striations and present our main conclusions in Sect.~\ref{sec:conclusions}. 
 
 \begin{figure*}[!htb]
       \centering
       \includegraphics[width=\hsize]{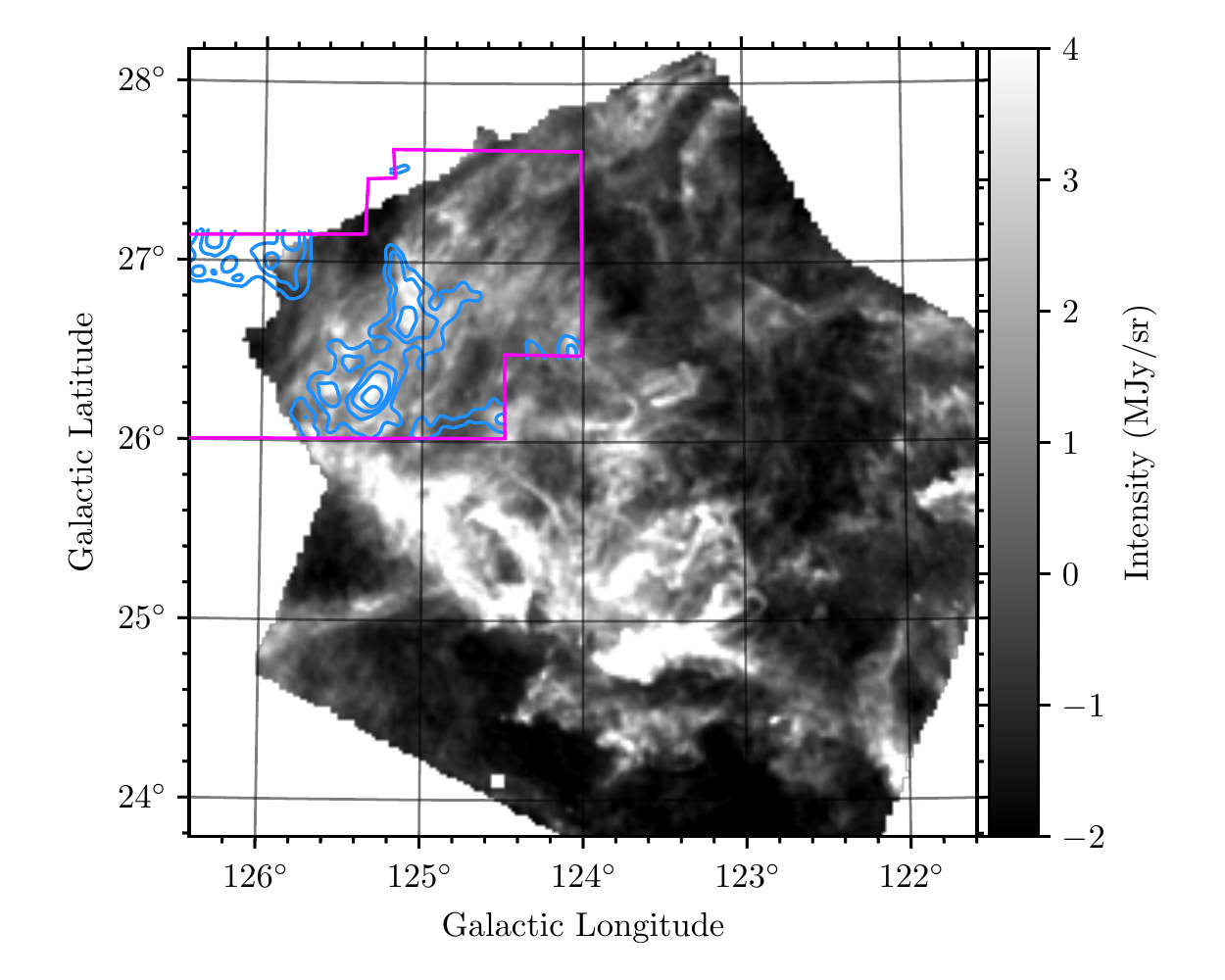}
       \caption{Large--scale view of the Polaris Flare, as observed with the Spire instrument of the {\it Herschel} observatory at 500 \mum. The magenta polygon shows the CO surveyed region, while blue contours show the smoothed integrated CO (J=2-1) intensity emission at 2.5 and 4 K~\kms.}
       \label{fig:polaris_large_scale}
   \end{figure*} 
   
    \begin{figure*}[!htb]
       \centering
       \includegraphics[width=\hsize]{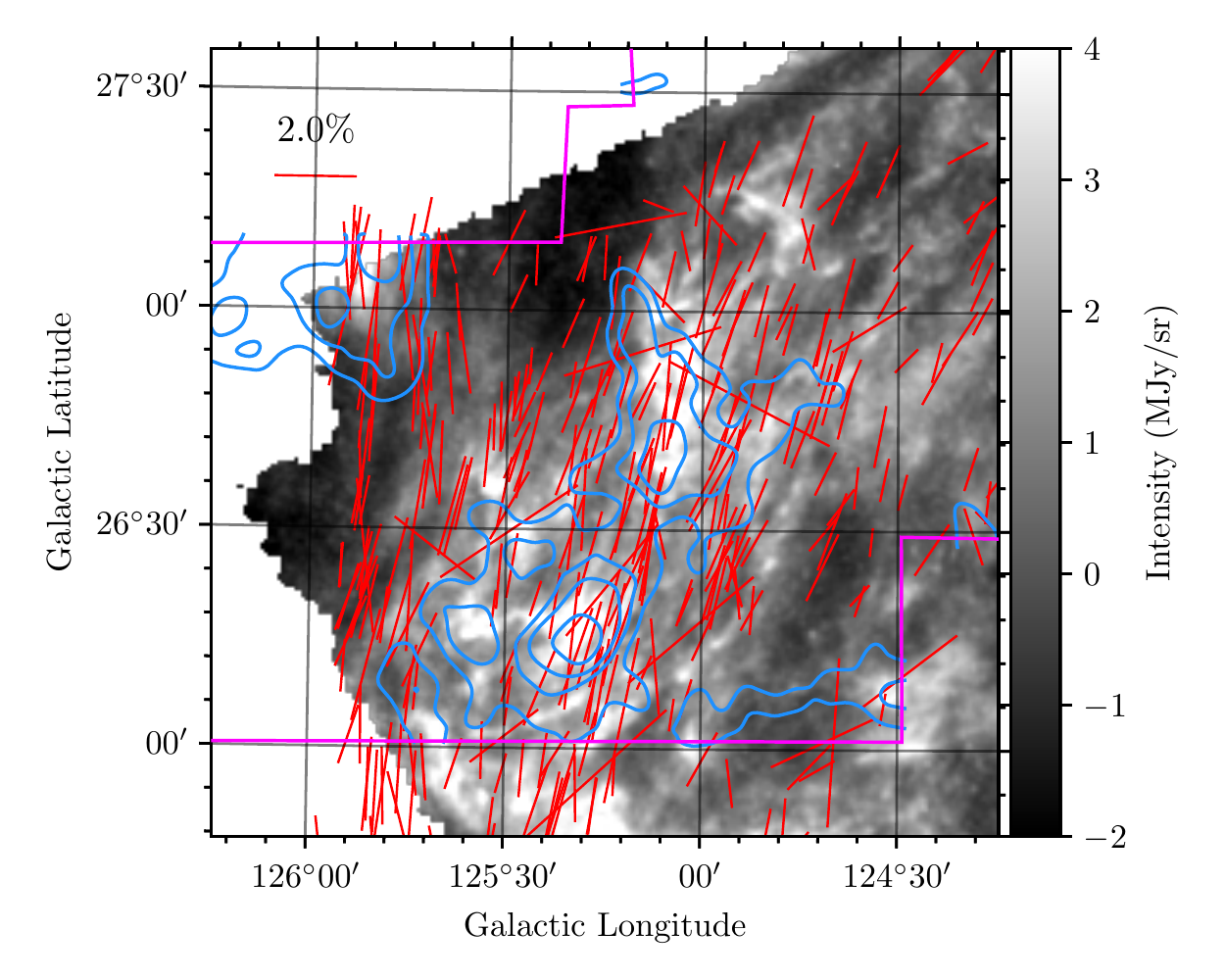}
       \caption{Zoomed--in view towards the striations region. Optical polarization segments (red lines) are overplotted on the dust continuum emission map of the Polaris Flare as observed with \textit{Herschel} at 500 $\mu$m. A line is shown for scale in the top left corner of the figure, corresponding to a degree of polarization of $p=2\%$. Blue contours correspond to the \CO\ integrated intensity as in Fig.~\ref{fig:polaris_large_scale}}.
    \label{fig:herschel_polaris_polarization}
   \end{figure*} 

\section{CO observations and data reduction}
\label{sec:co_observations}

We performed a survey of the \CO\ (J = 2$-$1) and \COThirteen\ (J = 2$-$1) transitions towards a region of the Polaris Flare that exhibits a pattern of striations in dust emission \citep{andre_2010}. The surveyed area covers $\sim 2.5$ square degrees and is indicated by the magenta polygon in Figs.~\ref{fig:polaris_large_scale} and~\ref{fig:herschel_polaris_polarization}; overlaid on the \textit{Herschel} dust emission map at 500~\mum, which has comparable spatial resolution with our CO survey.

Observations were conducted in January 2016 and between March and June 2017 with the 10--m Arizona Radio Observatory Heinrich Hertz Submillimeter Telescope (SMT). We used the dual-polarization ALMA Band 6 prototype sideband--separating mixer system, which allows simultaneous measurement of the \CO\ (230.5 GHz) and \COThirteen\ (220.4 GHz), J = 2 - 1, lines. The receivers were tuned so that the \CO\ (J = 2 - 1) line was measured in the upper sideband and the \COThirteen\ line in the lower sideband. The spectrometer used was a filterbank with 128 channels of 0.25 MHz bandwidth, corresponding to a velocity resolution of 0.325 km/s and 0.341 km/s at 230 GHz, and 220 GHz respectively. The velocity ranges of the spectra are [-23.6, 17.6] km/s for \CO\ and [-24.6, 18.6] km/s for \COThirteen. The beam size (FWHM) is 34".

The area was divided into 119 submaps of size 10\arcmin $\times$ 10\arcmin. Each map was scanned in On-The-Fly (OTF) mode. Our setup was identical to that described in \cite{Bieging_2014}. Scanning times varied according to atmospheric conditions, but were on average 2h per submap. The telescope scanned along lines of constant Galactic latitude in a boustrophedonic pattern at a scanning rate of 10\arcsec, 15\arcsec or 20\arcsec , depending on atmospheric conditions. Spectra were sampled every 0.1~s and later smoothed by 0.4~s as in \citet{Burleigh2013}. There were 60 rows scanned per submap, corresponding to Nyquist sampling perpendicular to the scanning direction. Telescope pointing was checked regularly. The total observing time was 377 hours. 

The receiver has two independent mixers, one for horizontal polarization and one for vertical. Since the CO line is unpolarized, the radiation from the molecular cloud is split equally between the two polarizations. We average both polarizations to improve the signal--to--noise ratio by $\sqrt 2$, assuming the mixers perform equally well, which was true for most of the observations. On some occasions the vertical polarization receiver had some problems, which resulted in highly-variable baselines. In those cases we retained only the data from the horizontal polarization receiver. 

The data of each submap were processed within the CLASS software. A flat baseline was subtracted from each spectrum and the data were convolved to a rectangular grid in Galactic coordinates. The resulting FITS files were combined to a single map via the function \texttt{XY\_MAP} of GILDAS. The data were put on the T$a^*$ scale and rescaled to main beam brightness temperature, $T_{mb}$, by dividing with the main beam efficiency of 0.8 (J. Bieging, priv. comm.). Due to the varying atmospheric conditions, the root mean square (rms) noise of the final map is not constant. The median of the distribution of the rms temperature $\rm T_{rms}$ is 0.47 K for the \CO\,  map.

\begin{figure*}
    \includegraphics[scale=1]{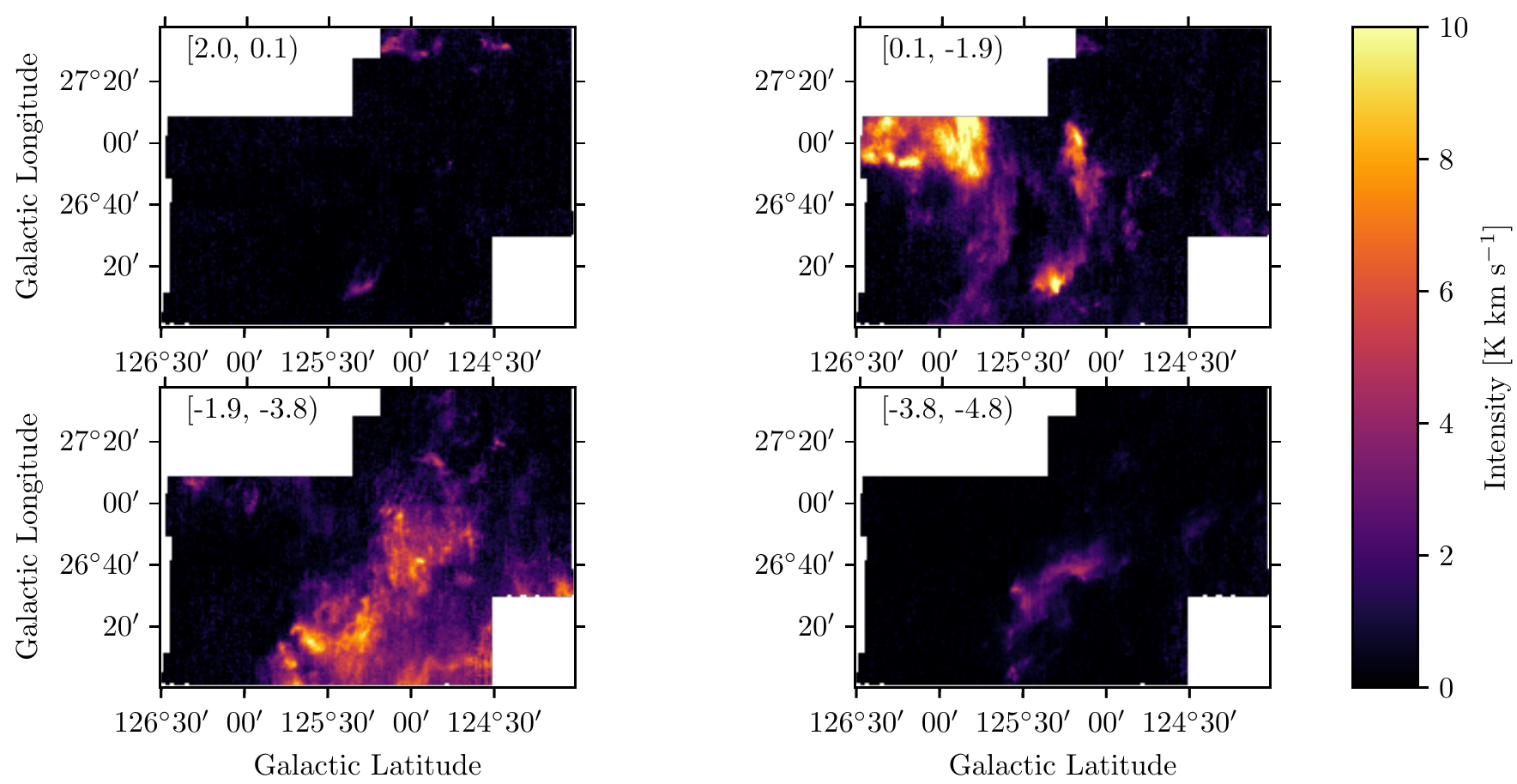}
    \caption{Integrated intensity maps of CO (J=2-1). The velocity integration interval is shown in the upper top left corner of each panel; velocities are in \kms.}
    \label{fig:CO_velocity_slices}
\end{figure*}
    
\section{Starlight polarization data}
\label{sec:polarization_data}

Optical polarimetric observations were performed by \cite{panopoulou_2015} with the RoboPol polarimeter \citep{king_2014, robopol_paper_2019} which is mounted at the Skinakas observatory in Crete, Greece\footnote{\url{http://skinakas.physics.uoc.gr}}. RoboPol is a quadruple-beam (four-channel) imaging polarimeter, which measures the $q$ and $u$ relative Stokes parameters simultaneously. This is achieved by splitting the observed light ray into four beams with two Wollaston prisms. For this work, we use stellar measurements in the entire unobstructed field of view (FOV) of RoboPol, as detailed in \cite{panopoulou_2015}. 

RoboPol is a one--shot polarimeter that can achieve accuracy similar to that obtained with conventional dual-beam polarimeters in significantly less time, due to the absence of moving parts. The major limitation of the design of RoboPol is that a single source is projected into four spots on the same CCD camera. As a result, the point spread functions of nearby stars can sometimes overlap on the CCD (we refer to these sources as `blended'). We have employed a supervised machine learning algorithm in order to remove blended measurements.

We apply several selection criteria to remove blended sources or sources affected by other systematics. In the first step, we excluded measurements with signal-to-noise ratio (S/N) in the degree of polarization less than 2.5. Such low S/N measurements introduce statistical biases in the degree of polarization \citep{vaillancourt_2006, Plaszczynski_2014}, and the spread of the the polarization angle distribution \citep{pattle_2017}, which is used in the estimation of the magnetic field strength (Sect.~\ref{sec:bpos_strength}). Then we removed overlapping stars using a Convolutional Neural Network (CNN). Finally, we used the empirical relation between the degree of polarization and dust extinction, in order to remove outliers from the sample. The last two steps are described in detail in Appendix~\ref{sec:CNN_model}.
   
     \begin{figure*}
       \centering
       \includegraphics[width=\hsize]{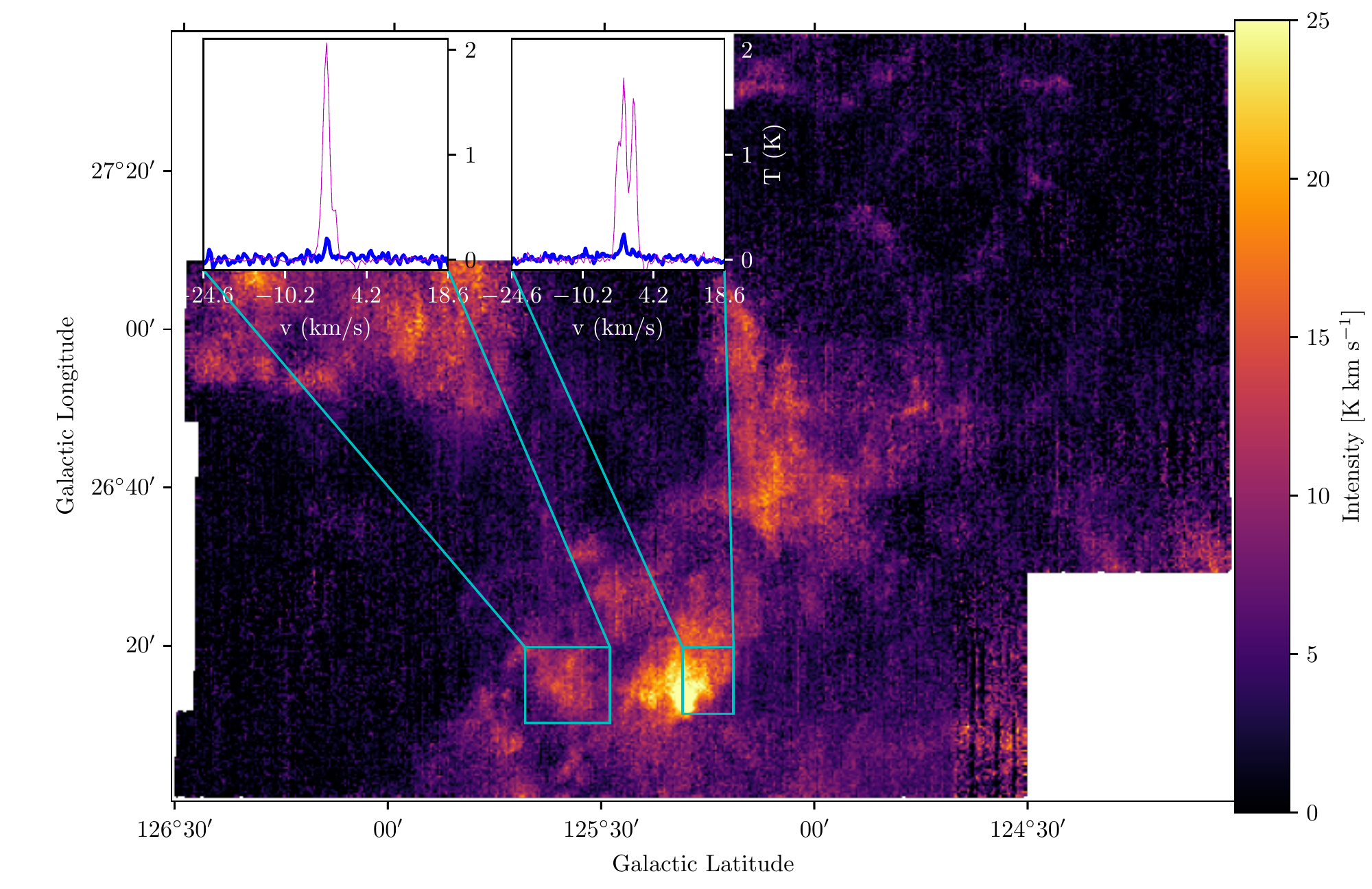}
       \caption{CO integrated intensity moment map at native resolution, without any filtering or masking. Cyan box--shaped regions 
       are defined by the positions where we detected the $^{13}$CO (J=2-1) line. Inset panels show the mean spectra 
       of $^{12}$CO (magenta), and $^{13}$CO (blue) within the cyan boxes. The intensity of the $^{13}$CO emission is approximately an order of magnitude lower than $^{12}$CO.}
       \label{fig:13CO}
   \end{figure*}
   
\section{CO gas properties}
\label{sec:co_striations}

\begin{figure*}
   \centering
    \includegraphics[width=\hsize]{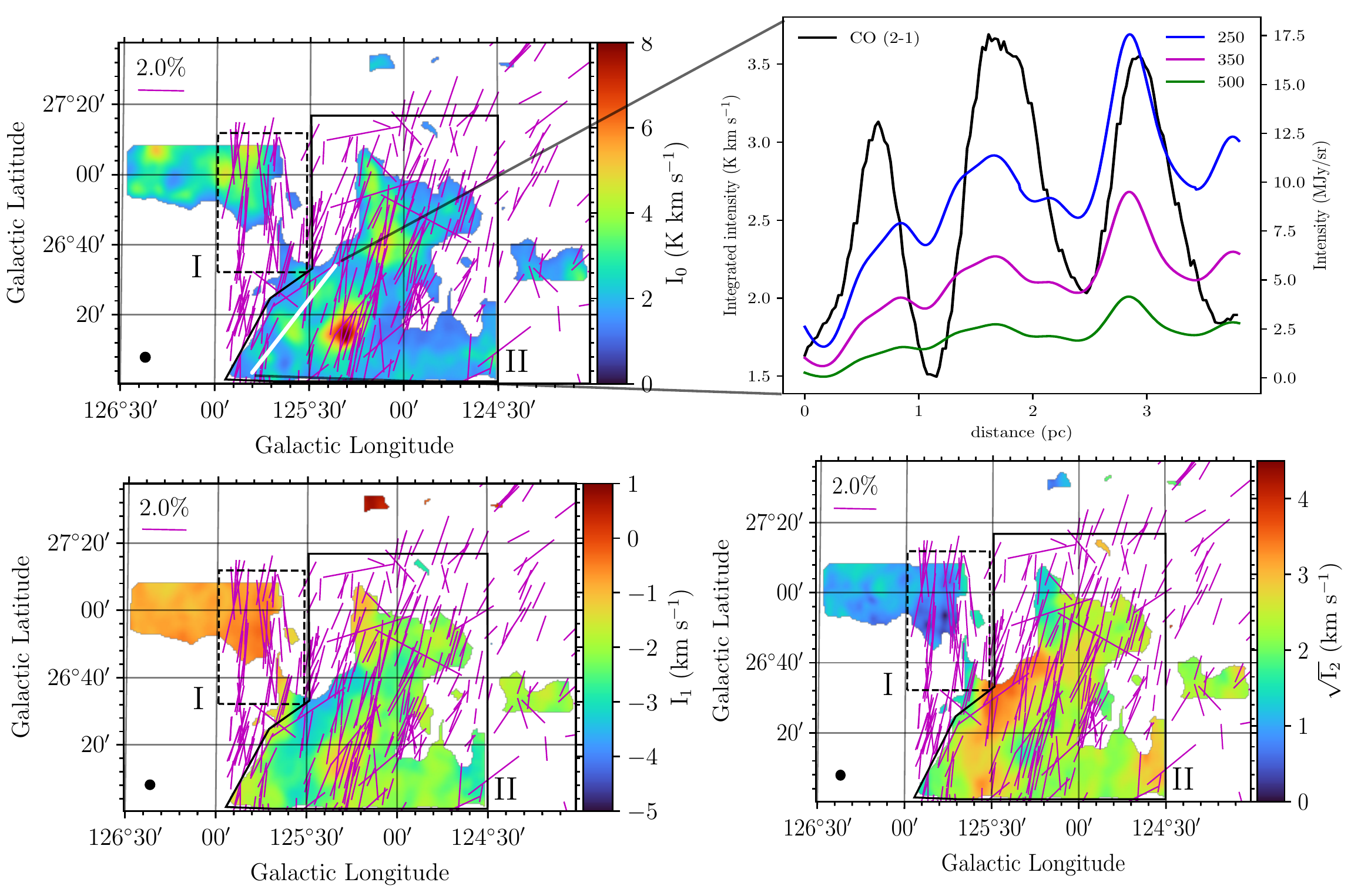}
    \caption{\CO\ (J=2-1) moment maps smoothed at a resolution of $2.57 \arcmin$. The circular beam in the left corner of each panel shows the Gaussian kernel used for smoothing the data. Magenta segments show the POS magnetic field orientation as traced by optical dust polarization. A scale segment is shown in the upper left corner in each panel. Pixels with total CO intensity less than 1.1 (K~\kms) have been masked out. The black dotted rectangle defines the boundaries of Region I, while the black solid polygon defines Region II. The black labels (I, II) are the region identifiers. \textbf{Top left panel:} Intensity (zero-moment) map of CO. \textbf{Upper right panel:} Intensity profiles along the white line shown in the $\rm I_{0}$ map. The left vertical axis corresponds to the CO (J=2-1) integrated intensity. The right vertical axis corresponds to the dust intensity of the three \textit{Herschel} bands of the SPIRE instrument. The wavelength of each band is shown in the legend. The quasi--periodicity of the CO intensity fluctuations are prominent across this line. The wavelength of the pattern is $\sim 1$ pc. Dust intensity profiles are more complex than the CO profile, showing several sub-parsec features. \textbf{Bottom left panel:} Centroid velocity (I$_{1}$, first moment) map of CO. \textbf{Right panel:} Velocity dispersion of CO ($\sqrt{\rm I_{2}}$, square root of the second moment map).}
   \label{fig:co_moment_maps}
\end{figure*}

The CO moments maps are constructed using the equation \citep{miville_2003, miville_2007}:
\begin{equation}
    \label{eq:momnet_maps}
    \rm I_{m} = \frac{\sum T_{mb} v^{m} \Delta v}{\rm I_{0}},
\end{equation}
where $\rm I_{0}$ is the zeroth moment map, defined as $\rm I_{0} = \sum T_{mb} \Delta v$, $\Delta v = 0.32$ \kms\ is the spectral resolution of the CO data, and $m = 1, 2$ are the indices of the first, and second moments respectively. The zeroth moment represents ($\rm I_{0}$) the total CO intensity measured in K~\kms, the first moment (I$_{1}$) is the LOS--averaged velocity measured in \kms, while the square of the the second moment ($\sqrt{\rm I_{2}}$) corresponds to the velocity dispersion measured in \kms.

We explore the CO morphological and kinematic properties in this Section. Fig.~\ref{fig:CO_velocity_slices} shows four different 
maps of the CO (J = 2 - 1) intensity integrated over various small velocity ranges, which are indicated in the upper left corner of each panel. Velocities are measured with respect to the Local Standard of Rest (LSR). 

The majority of CO gas has negative velocities, consistent with the CO (J=1-0) map of the region shown in \cite{Dame2001}, which had a beam size of $0.25$ square degrees. Our new CO (J=2-1) survey reveals the morphology of the gas at a detail comparable to that of \textit{Herschel}; the FWHM of the SPIRE beam at 500 \mum\ is $36.6 \arcsec$ \citep{griffin_2010}. In Fig.~\ref{fig:CO_velocity_slices}, the upper left panel, which corresponds to positive velocities, shows some sparsely--distributed CO structures but their emission is very weak and may not be associated with the Polaris Flare. We distinguish two negative velocity slices which show maximum CO emission. In the upper right panel, there is a bright region of CO 
in the upper left part of the surveyed region, at $l, b \sim 126\degr 00\arcmin, 27\degr 00\arcmin$ with an intensity of $\sim 10$ K~\kms. In the bottom left panel, the majority of CO is emitted from the lower right portion of the surveyed region towards $l, b \sim 125\degr 30\arcmin, 26\degr 20\arcmin$. There is also a CO structure in the bottom right panel, which corresponds to velocities from -3.8 to -4.8 \kms, but its intensity is weaker compared to other panels. The majority of CO in the striations region in the Polaris Flare moves with a LOS velocity in the range [0, -3.8] \kms. 

Fig.~\ref{fig:13CO} shows the $\rm I_{0}$ map of $^{12}$CO (J=2-1), which is integrated over the entire velocity range of the spectra. The map is noisy, mainly due to bad weather conditions during the observations, and shallow integration times needed for covering a wide area. We computed the amplitude of the noise ($\sigma_{\rm noise}$) by calculating the intensity rms in velocity channels free of CO emission. We find that $\sigma_{\rm noise}$ varies from 0.35 K~\kms\ up to $\sim 0.87$ K~\kms within the surveyed region; high noise is encountered close to the edges of this region. We consider the median $\sigma_{\rm noise}$, which is  $\sim 0.45$ K~\kms, as the characteristic amplitude of the noise in the CO data. 

The observed $\rm I_{0}$ map does not show the prominent CO linear structures found in regions with striations within other molecular clouds \citep[e.g.,][]{goldsmith_2008}. The observed CO structures have relatively low aspect ratios (blobby more than filamentary) (Fig.~\ref{fig:13CO}). We observe enhanced CO intensity at regions where the dust emission is maximum (Fig.~ \ref{fig:herschel_polaris_polarization}), but the morphologies of the observed CO and dust intensity structures are not well correlated. Dust structures are thin and aligned with the POS magnetic field, while CO structures are thick and inclined with respect to the POS magnetic field orientation. The dust emission maps in Figs.~\ref{fig:polaris_large_scale}, and~\ref{fig:herschel_polaris_polarization} correspond to the 500 \mum\ band of the SPIRE instrument on \textit{Herschel}. The morphology of the dust emission structures in the other two SPIRE bands at 200 and 350 \mum\ show the same behaviour.

In order to minimize the noise in the moment maps, we smoothed the data with a Gaussian kernel of FWHM = 2.5$\arcmin$; this is the minimum kernel size which minimizes sufficiently the noise in the map. After the smoothing we constructed the three moment maps. Then, we masked out every pixel with integrated intensity, $ \rm I_{0} \leq 2.5 \sigma_{\rm noise}$. In the top left panel of Fig.~\ref{fig:co_moment_maps}, we show the masked and smoothed $\rm I_{0}$ map.

In the $\rm I_{0}$ map, we observe a prominent quasi-periodic structure. In the top right panel of Fig.~\ref{fig:co_moment_maps}, we show the CO intensity profile along the white line in the $\rm I_{0}$ map (upper left panel in Fig.~\ref{fig:co_moment_maps}). The CO intensity as a function of distance along this line oscillates quasi--periodically with characteristic wavelength of the order of $\sim 1$ pc, assuming that the Polaris Flare is located at $\sim 340$ pc \citep{panopoulou_2022}. In the same panel, we also display the dust emission intensity as a function of position in the three SPIRE bands at 250, 350, and 500 \mum, obtained from maps smoothed to the same low resolution of the smoothed $\rm I_0$ map. The three \textit{Herschel} bands present almost identical morphological features except for the normalization of the intensity which varies due to the modified-black body nature of the dust emission. In the dust emission profiles, we observe a large-scale fluctuating pattern at $\sim 1$ pc, similar to that found in the CO emission. However, the first dust intensity local maximum at $\sim 1$ pc is not co--located with the CO peak. The dust intensity profiles present sub-parsec fluctuations, which are not observed in the CO intensity. The small-scale fluctuations maybe the reason why the dust intensity peak at $\sim 1$ pc does not match exactly with the CO intensity peak. The observed periodic pattern in the CO intensity is $17\degr$ off the mean POS magnetic field orientation, which is traced by dust polarization (magenta segments). In contrast to previous observations, where quasi-periodic CO structures tend to be orthogonal to the magnetic field orientation \citep{goldsmith_2008, palmeirim_2013_striations, cox_2016}, we observe quasi-periodic structures which tend to be parallel to the mean magnetic field orientation. We discuss noise and resolution effects in Sect.~\ref{sec:striations_CO_chemistry}.

We inspected the $\rm I_{1}$ map for potential evidence of quasi-periodicity similar to that one observed in the $\rm I_{0}$ map. However, correlating the $\rm I_{0}$ and $\rm I_{1}$ maps is challenging, because in some cases the quasi-periodic wave pattern may be imprinted in either the $\rm I_{0}$ or in the $\rm I_{1}$ map, but never in both maps. We present two thought experiments in order to demonstrate the above statement: a) If we consider that the quasi-periodicity is induced by a propagating plane wave polarized in the POS, then this wave should have no LOS velocity component, hence should be undetectable in the $\rm I_{1}$ map. b) If the polarization of the wave is along the LOS, then the wave should not be imprinted in the I$_{0}$ map. We can imagine multiple plane waves polarized along the LOS that induce a detectable pattern in the $\rm I_{1}$ map, but are undetectable in the $\rm I_{0}$ map. Therefore, we refrain from drawing any conclusion from the correlation of the $\rm I_{0}$ and $\rm I_{1}$ maps.

\begin{figure}
       \centering
       \includegraphics[width=\hsize]{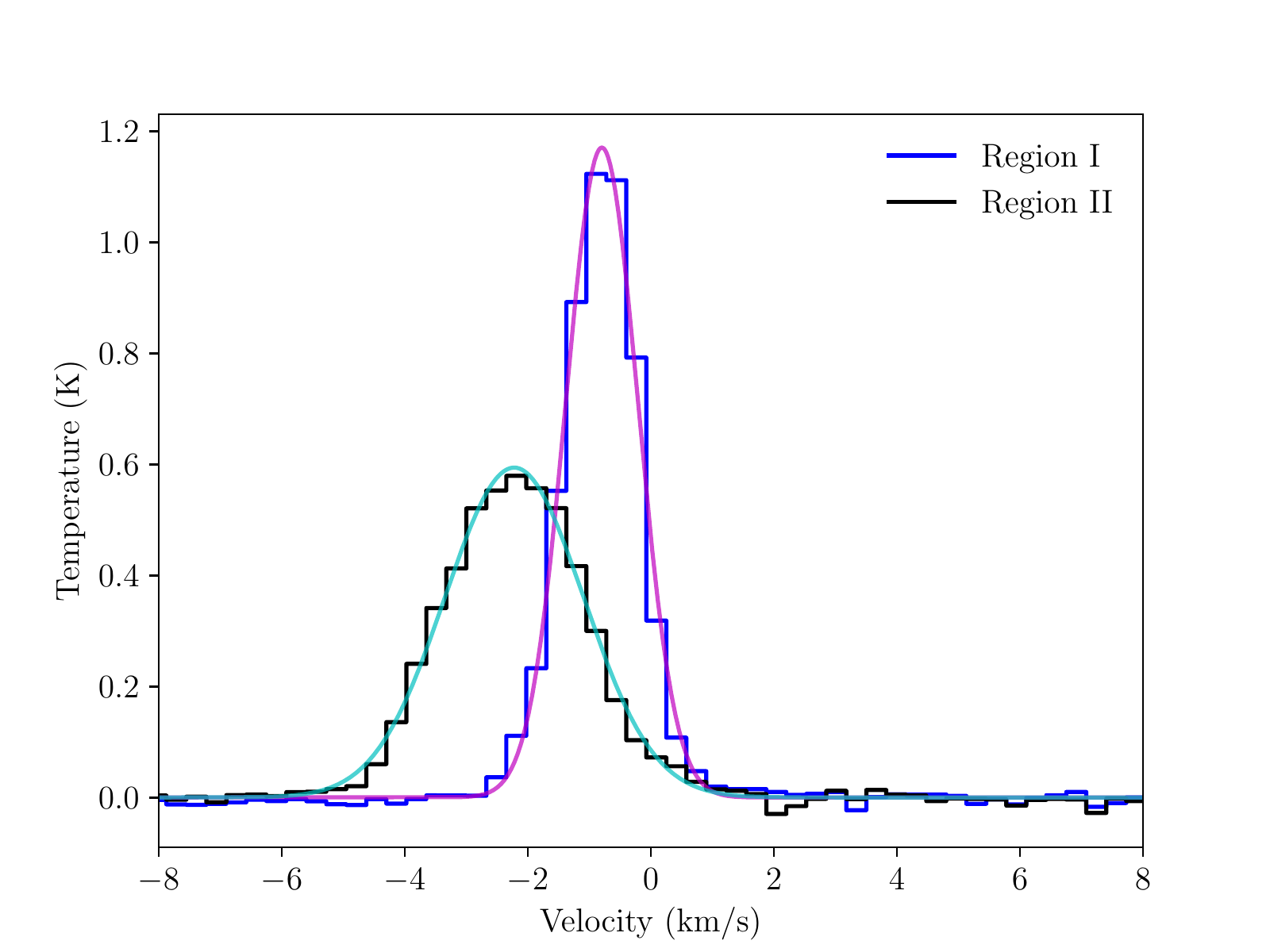}
       \caption{Averaged CO spectra towards Region I and II. The width of each bin is equal to the spectral resolution. Colored lines correspond to Gaussian fits. The mean velocities of Region I, and II are -0.80, and -2.22 \kms, respectively. The FWHM line widths in Region I and II are 1.39 and 2.61 \kms, respectively.}
       \label{fig:co_emission_spectra}
   \end{figure}

In the zeroth moment map, we identify two regions with prominent CO emission. The first region (Region I) is shown with the dotted, black rectangle in Fig.~\ref{fig:co_moment_maps}, while the second region (Region II) is shown with a solid, black polygon. These two regions have different kinematic properties. Region I has LSR velocities close to $0$ \kms, as indicated by the first moment map (bottom left panel in Fig.~\ref{fig:co_moment_maps}), and velocity dispersion close to $\sim 1$ \kms, as indicated by the square root of the second moment map (bottom right panel). Region II, has LSR velocities ranging from -4.0 to -1.0 \kms\ (bottom left panel of Fig.~\ref{fig:co_moment_maps}), and the velocity dispersion is on average close to 2 \kms, with maximum dispersion close to 4 \kms. In Fig.~\ref{fig:co_emission_spectra} we show the averaged CO spectra of the two regions.
   
We detected $^{13}$CO (J=2-1) only towards a few lines of sight, which coincide with peaks in the $^{12}$CO intensity. Cyan boxes in Fig.~\ref{fig:13CO} correspond to the regions where we detected the $^{13}$CO line. In these regions, the $^{13}$CO lines are $\sim 4 \sigma$ detections and their maximum intensity is almost an order of magnitude lower than that of $^{12}$CO (inset panels). 

\section{Magnetic field properties}
\label{sec:bpos_properties}

\subsection{Magnetic field morphology}
\label{subsec:bpos_morphology}

We explore the magnetic field morphology of Region I, and II. Magenta segments in Fig.~\ref{fig:co_moment_maps} correspond to the POS magnetic field orientation, as traced by optical dust polarization data. In Region I, the magnetic field orientation is on average parallel to the Galactic Latitude axis, while in Region II the magnetic field orientation is offset with respect to the Galactic Latitude axis. The difference in the mean magnetic field orientation of the two regions is more prominent in the distribution of polarization angles (Fig~\ref{fig:evpa_distributions}); polarization angles are reported in the Galactic reference frame, following the IAU convention. In Region I, the mean polarization angle is $-2\degr$, while in Region II, the mean angle is $-16\degr$.

Overall, the two regions have different CO properties (morphological and kinematic) and mean magnetic field orientations. From the 3D extinction map of \cite{green_2018}, we found that the two regions are not co--located. Region II is located at $\sim 335$ pc, while Region II at $\sim 340$ pc. Both regions could be parts of the Polaris Flare, but the 3D extinction map is noisy there, and we cannot make stronger conclusions. We also explored the 3D map of \cite{leike_2020}, but it is also noise--dominated towards this region. We do not observe any sign of interaction between the two regions. 

\subsection{Magnetic field strength}
\label{sec:bpos_strength}

We estimate the POS magnetic field strength (\Bpos) of each region following the method of \cite{skalidis_tassis_2021}. The POS magnetic field strength can be estimated using: 
\begin{equation}    
    \label{eq:st_bpos}
    \rm B_{POS} \approx \sqrt{4 \pi \rho}\frac{\sigma_{u, turb}}{\sqrt{2 \delta \theta}},
\end{equation}
where $\rho$ is the gas density, $\sigma_{\rm u, turb}$ is the gas turbulent velocity, and \dtheta\ the dispersion of polarization angles. 

The assumption of this method is that gas turbulent kinetic energy is exchanged between kinetic and magnetic forms. Eq.~\ref{eq:st_bpos} has been tested in ideal-MHD simulations with no self--gravity, and was found to produce accurate estimates of the projected mean magnetic field strength for a wide range of Alfvénic ($0.1 \leq $~\MA~$\leq 2$), and sonic ($0.5 \leq$~\MS~$\leq 20$) Mach numbers \citep{skalidis_2021_Bpos}.

\subsubsection{Polarization angle intrinsic spread}
\label{subsec:intrnsic_spread}

    \begin{figure}
       \centering
       \includegraphics[width=\hsize]{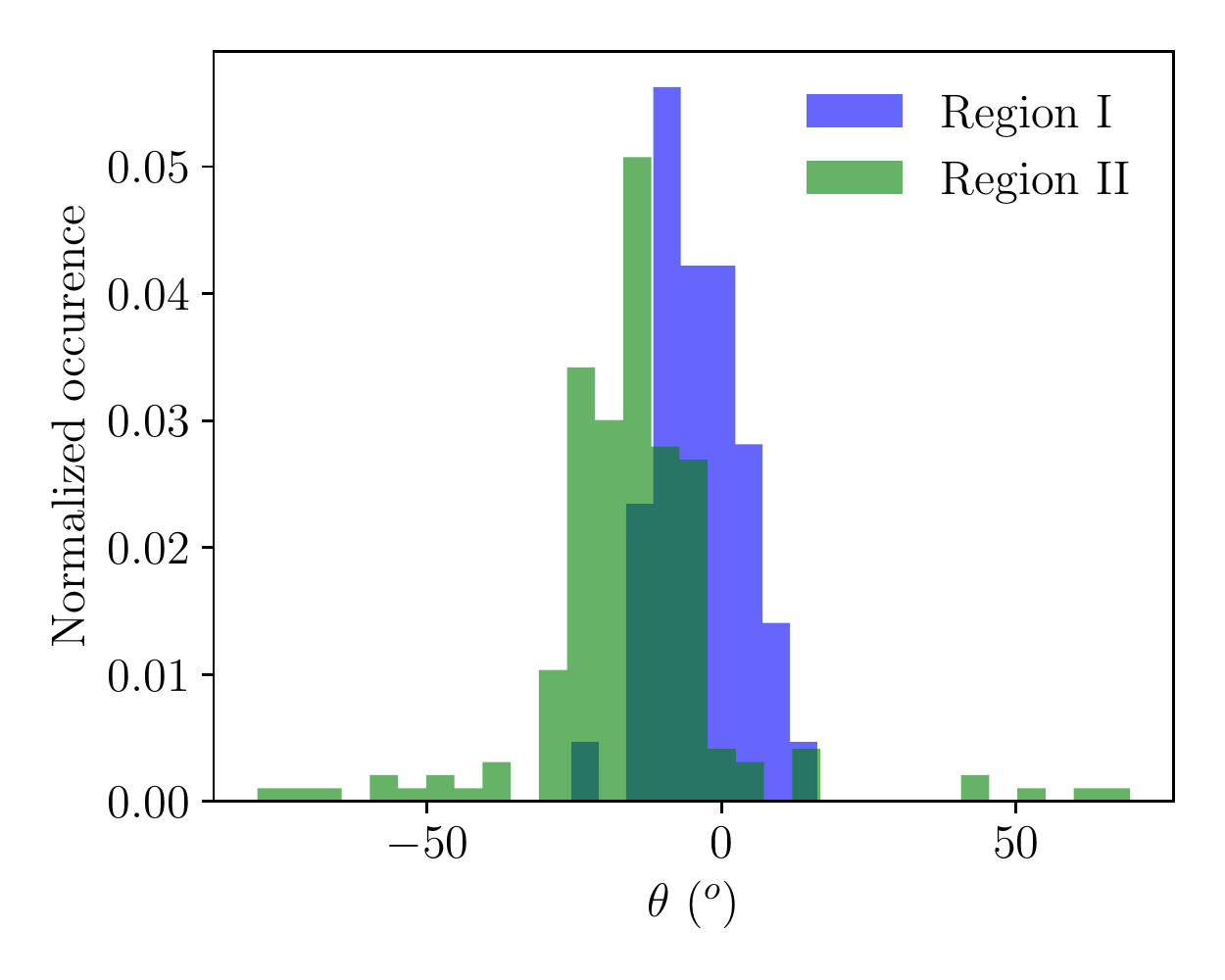}
       \caption{Normalized distributions of polarization angles for both regions. The size of each bin has been optimized using Bayesian analysis from \textit{astropy}, in order to reflect the intrinsic shape of each distribution.}
       \label{fig:evpa_distributions}
    \end{figure}

   \begin{figure*}
       \centering
       \includegraphics[width=\hsize]{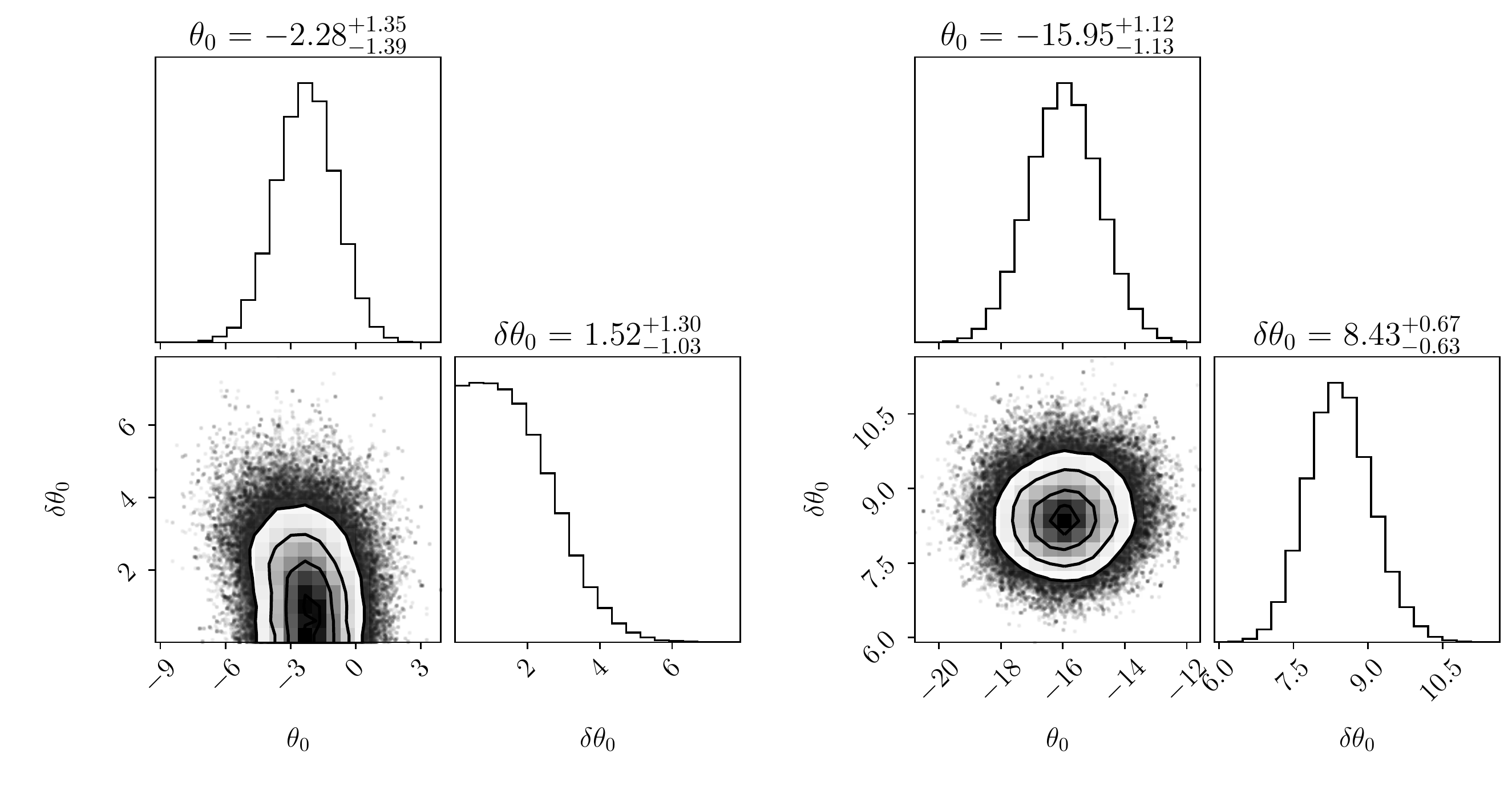}
       \caption{1D, and 2D marginalized posterior distributions of $\theta_{0}$, and $\delta \theta_{0}$ for Region I (left panel), and II (right panel). These plots were produced with the corner plot function \cite{corner}.}
       \label{fig:evpa_posterior}
   \end{figure*}

In the \Bpos\ estimation (Eq.~\ref{eq:st_bpos}), \dtheta\ corresponds to the intrinsic spread induced by gas turbulent motions. Estimating the intrinsic dispersion of polarization angles (\dthetaIntrinsic) from the observed distribution of the polarization angles may be challenging when S/N $\lesssim 3$. Then the observational uncertainties of the polarization angles are comparable to the observed spread of the polarization angle distribution. Observational uncertainties tend to bias the estimation of \dthetaIntrinsic\ towards larger values \citep{pattle_2017}; overestimating \dthetaIntrinsic\ leads to underestimating \Bpos\ (Eq.~\ref{eq:st_bpos}). In Regions I and II the maximum observational uncertainty of the polarization angles is $\sim 16\degr$, while the observed \dtheta\ are $\sim 7.7 \degr$ and $\sim 15 \degr$ respectively (Fig.~\ref{fig:evpa_distributions}). Thus, the observed \dtheta\ are likely biased due to the large observational uncertainties. The statistical bias in \dtheta\ is usually treated by subtracting the observed spread with the mean observational uncertainty in quadrature \citep[e.g.,][]{panopoulou_2016, skalidis_2022}. Here we employ a Bayesian approach in order to properly mitigate the observational uncertainties and derive the posterior distribution of \dthetaIntrinsic. 

We fit Gaussians to the polarization angle distributions of both regions, since numerical simulations show that this is the intrinsic shape of the polarization angle distributions induced by MHD turbulence \citep[e.g.][]{heitsch_2001, falceta_2008, skalidis_tassis_2021}. A normalized Gaussian distribution is characterized by two free parameters: 1) the mean angle ($\theta_{0}$), and 2) the spread, which is the intrinsic spread (\dthetaIntrinsic). We assumed uniform priors with $\theta_{0}~\epsilon~[-90\degr, 90\degr]$, and $\delta \theta_{0}~\epsilon~[0\degr, 50\degr]$; spreads larger than $50\degr$ are uninformative of the intrinsic (turbulent) spread due to the $\pi$ ambiguity in polarization angles. We also assumed Gaussian errorbars. Each polarization angle data point ($\theta_{i}$) is characterized by the following data equation,
\begin{equation}
    \theta_i = \mathcal{G}(\theta_0,\delta\theta_0) + \mathcal{G}(0,\sigma_i),
\end{equation}
where $\mathcal{G}$ denotes a Gaussian probability density function, and $\sigma_{i}$ is the observational uncertainty of the polarization angles. A detailed discussion about the Bayesian modelling of the intrinsic spread can be found in \cite{pelgrims_2022}.

We maximized the following log-likelihood using the "emcee" Markov Chain sampler \citep{emcee}, 
\begin{equation}
    \label{eq:log_likelihood}
    \ln{\mathcal{L}} = - \sum_{i=1}^{N} \left [ \ln{\sqrt{2 \pi}} + \ln{\delta \theta_{tot, i}} + \frac{(\theta_{i} - \theta_{0})^{2}}{2\delta \theta_{tot, i}^{2}} \right ],
\end{equation}
where $\delta \theta_{tot, i}^{2} = \delta \theta_{0}^{2} + \sigma_{i}^{2}$, and denotes the observed polarization angle spread. In the distribution of polarization angles of Region II, we applied a $4~\sigma$ clipping in order to remove measurements which are likely intrinsically polarized stars. The polarization signal of these stars  probes the properties of the intrinsic polarization mechanism and not the ISM magnetic field properties. In total only two measurements were discarded. 

The posterior distributions of $\theta_{0}$, and \dthetaIntrinsic\ for both regions are shown in Fig.~\ref{fig:evpa_posterior}. We computed the mean $\theta_{0}$, and $\delta \theta_{0}$ of the posterior distributions with their corresponding $68 \%$ confidence intervals. The $68 \%$ percentile for Region I is $\delta \theta_{0} = 2.28^{+1.30}_{-1.03}$, and for Region II is $\delta \theta_{0} = 8.4 ^{+0.67}_{-0.63}$ degrees. These values are significantly different than the intrinsic spread one would obtain by following the simplistic quadratic subtraction, which assumes uniform observational uncertainties, of the polarization angle spread and the mean observational uncertainty \citep[e.g.,][]{panopoulou_2016, skalidis_2022}; with the quadrature subtraction, we obtain for Region I and II that $\delta \theta_{0}$ is 1.87$\degr$ and 12.6$\degr$, respectively. In contrast, the Bayesian approach propagates the uncertainty of each measurement to the posterior distribution, and hence resembles better the intrinsic spread.

In Region I, there are only 46 reliable polarization measurements and $\delta \theta_{0}$ is an upper limit, which yields a lower limit for \Bpos. The intrinsic spread is statistically consistent with zero, which means that \Bpos\ estimated from equation \ref{eq:st_bpos} would be infinite. Since the upper limit of \Bpos\ is unconstrained there, we proceed with the magnetic field strength estimation only for Region II, where we have 202 trustworthy measurements and $\delta \theta_{0}$ is statistically constrained. 

\subsubsection{Density and turbulent gas velocity estimation}
\label{subsec:dens_estimation}

The gas density is $\rho = \mu m_{\rm{H_{2}}} n_{\rm{H_{2}}}$, where $\mu=2.8$ is the mean molecular weight, considering the helium abundance \citep[Appendix A1,][]{kauffmann_2008}. Quantities $m_{\rm{H_{2}}}$, and \nHmol\ denote the mass of the hydrogen molecule 
and the molecular hydrogen number density, respectively. 

In principle, one could use 3D extinction maps \citep[e.g.,][]{green_2018, leike_2020, lallement_2019} to estimate the LOS dimension of the Polaris Flare and hence \nHmol, as has been done for other clouds \citep{zucker_2021, tritsis_2022}. However, we found that the 3D extinction maps of \cite{green_2018} and \cite{leike_2020}, did not have enough sensitivity towards the striations region to draw any conclusion on the depth of the cloud. 

The mean density of the striations region has been previously considered to be $300 \leq$~\nHmol~$\leq 1000$ \VolDens\ \citep{panopoulou_2016}. However, by modelling the \CO, and \COThirteen\ emission lines, we infer that the maximum gas density can be as high as 1200 \VolDens\ (Appendix~\ref{sec:co_ratio}). In order to be conservative, we adopt this value as an upper limit for \nHmol, and hence we adopt that $300 \leq$~\nHmol~$\leq 1200$ \VolDens.

For the estimation of the turbulent velocities we use the linewidth of the CO emission spectra. The width of the emission lines is broadened due to thermal and turbulent gas motions. In order to estimate the turbulent broadening we subtract the thermal broadening in quadrature. The thermal broadening depends on the gas kinetic temperature (\TK), which is unknown. We assume \TK\ values based on the literature. \cite{heiles_troland1} studied the properties of \HI\ clouds in our Galaxy using \HI\ absorption data. They found that the mean temperature of atomic gas is $\sim 50$ K. Since the Polaris Flare is a diffuse molecular cloud, and gas in molecular clouds is colder than in atomic clouds, we consider \TK\ $= 50$ K to be an upper limit to the kinetic temperature. \TK\ of molecular gas is $\sim 1.5$ times smaller than that of atomic gas \citep{goldsmith_2013}, hence we adopt \TK~= 35 K as a reference temperature. We consider \TK\ = 20 K as the lower limit to the kinetic temperature. 

The observed spread of the CO (J=2-1) emission line is 1.13 \kms. For \TK\ = 50 K, and \TK\ = 20 K we estimate, by subtracting in quadrature the thermal line spread from the observed one, that the turbulent broadening is 1.03, and 1.09 \kms, respectively. Variations of \TK\ lead to a 0.06 \kms\ uncertainty in the estimated $\sigma_{\rm u, turb}$. If we consider that \TK\ = 100 K, which is a typical temperature of molecular gas in photo-dissociation regions (PDRs), then $\sigma_{\rm u, turb} \approx 0.92 $ \kms. This means that if we allow \TK\ to range from 20 to 100 K, the uncertainty on $\sigma_{\rm u, turb}$ becomes 0.2 \kms. However, for the striations region we favor \TK\ $\lesssim 50$ K (Appendix~\ref{sec:co_ratio}). In order to be conservative we consider that the uncertainty of $\sigma_{\rm u, turb}$ originating from \TK, is 0.1 \kms. Thus, we estimate the turbulent broadening to be $\sigma_{\rm u, turb} = 1.0 \pm 0.1$ \kms. We estimate that \Bpos\ (Eq.~\ref{eq:st_bpos}) ranges from 38 to 76 \muG, while in the same table we also report the \Bpos\ estimate at intermediate densities, n~=~600~\VolDens, for reference. The Alfv\'en speed, defined as \VA~=~\Bpos\ $/ \sqrt{4 \pi \rho}$, is \VA\ = $\sigma_{\rm u, turb}/\sqrt{2\delta \theta} = 2.0 \pm 0.1$ \kms.

\subsection{Sonic and Alfv\'enic Mach numbers}

According to the ST method, the projected Alfvénic Mach number is $\mathcal{M}_{A}^{\rm 1D} \equiv \rm {\sigma_{turb}/\rm V_{A}} \approx \sqrt{2 \delta \theta}$. We estimate the total \MA\ of the cloud by multiplying $\mathcal{M}_{A}^{\rm 1D}$ by $\sqrt{3}$, assuming isotropic turbulent properties. For Region II, we obtain that \MA~ $\approx 0.94 \pm 0.02$, which means that turbulence is sub-Alfvénic, hence the magnetic field is dynamically dominant there. The estimated \MA\ should be treated as an upper limit, because turbulence in the ISM is not isotropic \citep{higdon_1984, desai_1994, backer_2000, Shaikh_2007, planck_collaboration_2016}, and for anisotropic turbulence the factor which connects $\mathcal{M}_{A}^{\rm 1D}$ with the total \MA\ is less than $\sqrt{3}$ \citep{beattie_2020_Bfluctuations}. However, even for isotropic turbulence the correction factor which connects the projected with the 3D Mach numbers can be less than $\sqrt{3}$ \citep{stewart_2022}.
 
We estimate the sonic Mach number $\mathcal{M}_{s}^{\rm 1D} \equiv \rm \sigma_{turb}/\rm c_{s}$, where $\rm c_{s}$ is the sound speed. For isotropic turbulence, \MS\ $\approx \sqrt{3} \mathcal{M}_{s}^{\rm 1D}$ which yields that $\mathcal{M}_{s} \sim 3.9 - 6.5$ for Region II. All the estimated quantities are summarized in Table~\ref{table:B_properties}.

\begin{table*}[!htb]
\caption{Region parameters}             
\label{table:B_properties}      
\centering                          
\begin{tabular}{c c c c c c c c c c}        
\hline\hline                 
Region & $\theta_{0}$ ($\degr$) & $\delta \theta_{0}$ ($\degr$) & \MA & \TK\ & \MS & $\sigma_{turb}$ (\kms) & \VA  (\kms)  & $n$ (\VolDens)  &\Bpos (\muG)  \\    
\hline                        
\hline
  &                &               &              &  &  &  &  & 300 & 38 \\  
II & -16.0 $\pm$ 1.1 & 8.4 $\pm$ 0.7 & $0.94 \pm 0.02$ & 20 - 50 & $3.9 - 6.5 $ & $1.0 \pm 0.1$ & $2.0 \pm 0.1$  & 600 & 54 \\
  &                &               &                &  &  &  &  & 1200 & 76 \\  
\hline                                   
\end{tabular}
\end{table*}

\section{CO chemistry in striations and the role of magnetic fields}
\label{sec:striations_CO_chemistry}

Magnetic fields affect the dynamical properties of clouds in the ISM, and hence their chemical evolution. Dynamically important magnetic fields tend to suppress molecule formation  \citep{walch_2015, Pardi2017TheISM, seifried_2016, Girichidis2018TheClouds}. Magnetic fields are not easily compressed perpendicular to the field lines, due to magnetic tension and pressure. Simulated magnetized clouds are more diffuse than their corresponding non--magnetized counterparts. As a result, gas in magnetized clouds is not well shielded, hence magnetized clouds tend to have higher CO-dark \Hmolecular\ abundance than non--magnetized clouds, which are richer in CO \citep{Seifried2021}. 

Molecule formation is also affected by the relative alignment between the magnetic field and the cloud structure, or the velocity flow. For example, in the simulations of \cite{seifried_2016}, when the magnetic field is aligned with the main filament axis, the abundances of C, \CII, and CO change abruptly as we move from the outer parts to the center of the filament. When the magnetic field is perpendicular to the filament axis, the carbon-based species abundances have shallower gradients towards the center of the filament. Furthermore, amplified magnetic fields from colliding flows form a critical angle with the upstream velocity, which determines whether molecule formation will be suppressed or enhanced \citep{Iwasaki2019TheGases, Iwasaki2022UniversalGases}. Overall, all the aforementioned numerical studies suggest that the ISM magnetic field affects the formation of molecular gas, mainly by setting a preferred orientation along which gas accumulation, and hence molecule formation, can take place. Observationally, there is evidence supporting this theoretical scenario \citep{skalidis_2022}.

In the current study, we explore the role of magnetic fields in the molecule formation in striations. Compressions in our target region are most likely formed by propagating MHD waves \citep{tritsis_2016}, as indicated by the prominent periodic structure in the CO intensity map (Fig.~\ref{fig:co_moment_maps}), and not by colliding flows, as is the case of the aforementioned numerical simulations. Therefore, we cannot make a direct comparison of our observations and the simulated colliding flow models. However, \VA\ seems to be an important parameter for initiating the suprathermal chemistry \citep{federman_1996, visser_2009}. Below we explore the conditions under which wave modes could contribute to the CO formation within the striations region in the Polaris Flare. 

\subsection{Compressible MHD modes as agents of CO formation}

Striations are considered to form due to the propagation of compressible MHD modes \citep{tritsis_2016}. Strong observational evidence supporting this theory comes from the quasi-periodic patterns observed in CO, and dust intensity maps \citep{goldsmith_2008, palmeirim_2013_striations, cox_2016, malinen_2014_striations}. In the Polaris Flare, we find a prominent quasi--periodic pattern in the CO zeroth moment and dust intensity maps, with periodicity almost parallel to the mean magnetic field orientation (Fig.~\ref{fig:co_moment_maps}), and not perpendicular as has been previously observed \citep{goldsmith_2008}. However, we note that the dust emission map is periodic both along and across the mean magnetic field orientation.

The contrast in the CO zeroth moment map ($\sim 43 \%$) is larger than the contrast of the \NHmol\ map ($18 \%$). Differences in the column density contrast between gas and CO maps have also been observed in numerical simulations \citep{tritsis_2016}. The difference in the contrast of the two maps indicates that it is unlikely that CO could have pre--existed and been compressed by a propagating MHD wave. It is more probable that the CO formation has been enhanced by a propagating MHD wave. Firstly, we examine the case of waves in compressible gas, and then the incompressible case. 

Compressible linear MHD perturbations can propagate in the form of fast, and slow magnetosonic modes. When \VA~$\gg c_{s}$ the characteristic propagation speed for the fast and slow modes is \VA, and $c_{s}$ respectively. Fast modes propagate isotropically, while slow modes, which oscillate as sound waves when \VA~$\gg c_{s}$, cannot propagate perpendicular to the magnetic field lines. Thus, both slow and fast modes could form the quasi-periodic pattern in the observed CO zeroth-moment map (Fig.~\ref{fig:co_moment_maps}), since the propagation axis of this pattern is slightly off from the mean magnetic field orientation. 

The characteristic timescale of CO formation (\tco) in general differs from the MHD wave periods ($\tau_{\rm waves}$). If \tco\ > $\tau_{\rm waves}$, then MHD waves compress and rarefy the gas before CO has enough time to form. If \tco\ < $\tau_{\rm waves}$, then CO can form at an enhanced rate within the  compressed regions produced by the passage of the MHD wave. The slow and fast modes have different characteristic timescales for the same wavelengths. Using the wavelength ($\sim 1$ pc) of the observed quasi--periodic pattern in the CO intensity map, we obtain that the period of a fast mode should be \tfast~$\approx 0.48$ Myr, while for a slow mode the period should be \tslow~$ = 2.1 - 3.4$ Myr. The estimated range of \tslow\ corresponds to the assumed \TK\ range (20 - 50 K). For the Polaris Flare, we estimate that \tco\ = 0.03 - 1.0 Myr (Appendix~\ref{app:co_formation_timescale}). If \tco\ = 1.0 Myr, then \tfast~$<$~\tco~$<$~\tslow. In this case, fast modes oscillate much faster than the typical CO formation timescale, hence are unlikely to have contributed to the CO formation. If \tco\ = 0.03 Myr, then \tco~$<$~\tfast~$<$~\tslow, hence both fast and slow modes could be responsible for the CO enhancement. We note that slow modes are slower than the CO formation timescale for both \tco\ limits.

In the Polaris Flare, the dust intensity maps follow a power-law behavior with spectral index -2.65 \citep{miville_2010_polaris_PS}. The inertial range of the power spectrum extends down to 0.01 pc, implying that MHD wave-modes with such short wavelengths could be observed. However, there is a minimum wavelength ($\rm \lambda_{f, min}$) above which fast modes could be observable in the CO (J=2-1) intensity map. The minimum wavelength corresponds to fast modes oscillating with periods \tfast\ 
$\geq$ \tco, where \tfast\ = $\rm \lambda /$ \VA. The condition \tfast\ $\geq$ \tco\ is satisfied when $\lambda \geq \rm \lambda_{f, min}$, where $\rm \lambda_{f, min} = \tau_{\rm CO} \times \rm V_{A} \approx 2$ pc for \tco\ = 1 Myr, and $\lambda_{\rm f, min} = 0.06$ pc for \tco\ = 0.03 Myr. 

For the slow modes, we estimate that the minimum wavelength is $\rm \lambda_{s, min} = \tau_{\rm CO} \times c_{s} \approx 0.28 - 0.44$ pc, when \tco\ = 1 Myr, where the uncertainty on $\rm \lambda_{s, min}$ comes from the \TK\ uncertainty (Sect.~\ref{subsec:intrnsic_spread}). Similarly, we estimate that the minimum wavelength for \tco\ = 0.03 Myr is  $\rm \lambda_{s, min} = \tau_{\rm CO} \times c_{s} \approx 0.008 - 0.013$ pc. Overall, slow modes are characterized by lower minimum wavelengths than fast modes due to their lower propagation speeds.

The CO intensity profile in Fig.~\ref{fig:co_moment_maps} does not show evidence of wave-modes with wavelengths shorter than 1 pc. However, this could be related to sensitivity issues, because sub--parsec wavelength modes might induce intensity contrast below the sensitivity limit of our observations. Thus, we cannot exclude the possibility of sub-parsec--wavelength modes enhancing the formation of CO, but the dominant detected wavelength is $\sim 1$ pc. The sub--parsec wave features in the dust emission profiles (upper right panel in Fig.~\ref{fig:co_moment_maps}) are close to the spatial resolution limit, hence might be artifacts of the smoothing.

\subsection{The role of Alfv\'en waves in the formation of CO}

Ambipolar diffusion decouples ions and neutrals, hence perturbations with timescales longer than the characteristic timescale of ambipolar diffusion do not propagate to neutral gas because they are damped. There are two types of ambipolar diffusion: 1) magnetically-driven, and 2) gravitationally-driven ambipolar diffusion \citep{mouschovias_2011}. Gravitationally-driven ambipolar diffusion requires the existence of a local center of gravity \citep[e.g., a central core which preferentially drags neutral particles past the ions increasing the mass--to--magnetic--flux ratio of cores relative to their surroundings,][]{mouschovias_2006, tassis_2014}. Magnetically--driven ambipolar diffusion operates at shorter timescales than the gravitationally--driven, which is usually slow since it operates at timescales equal to  $\sim 10$ Myr \citep{tassis_2004}. For this reason, we focus only on magnetically--driven ambipolar diffusion.

\cite{federman_1996} invoked the magnetically--driven ampibolar diffusion to explain the enhancement of ionic products, such as $\rm CH^{+}$, compared to neutral $\rm OH$. Due to the relative drift between ions and neutrals, Alfv\'en waves add more energy to \CII\ reactions than to oxygen--based reactions. Thus, the reaction $\rm C^{+} + H_{2} \rightarrow CH^{+} + H$ produces more $\rm CH^{+}$ than the $\rm OH$ produced by the reaction $\rm O + H_{2} \rightarrow OH + H$. This explains the observed overproduction of $\rm CH^{+}$, without producing much $\rm OH$. The enhanced $\rm CH^{+}$ abundance increases the formation rate of CO.  Alfv\'en waves propagate along the magnetic field lines with speed equal to \VA, which implies that Alfv\'en waves could locally enhance $\rm CH^{+}$, and hence CO, when their velocity is maximum; when the velocity is minimum, CO formation should be negligible. Thus, Alfv\'en waves could induce the observed quasi-periodic feature in the CO zeroth moment map (Fig.~\ref{fig:co_moment_maps}). However, we exclude Alfv\'en waves because the observed quasi-periodic CO profile is correlated with dust intensity (Fig.~\ref{fig:co_moment_maps}); Alfv\'en waves are incompressible and cannot induce temperature (or equally density) fluctuations.

\subsection{Uncertainties in the estimated wave periods}

\subsubsection{Uncertainties on \tfast}

The wave periods of fast, and slow magnetosonic modes depend on their propagation speed. The uncertainties of the propagation speeds, \VA\ for fast modes, and $\rm c_{s}$ for slow modes, propagate to the estimated wave periods, \tfast, and \tslow. \VA\ was constrained with high accuracy based on the dispersion of polarization angles with the method of \cite{skalidis_tassis_2021}. The major uncertainty of \VA\ comes from the assumptions of the aforementioned method. The \cite{skalidis_tassis_2021} method has been tested against a wide range of numerical simulations and its accuracy was found to be better than a factor of two \citep{skalidis_2021_Bpos}. 

Alternative methods for estimating \VA\ \citep[DCF,][]{davis_1951, chandra_fermi} tend to produce higher values than the method of \cite{skalidis_tassis_2021}. According to DCF, the Alfv\'en speed is $V_{A} = f~\sigma_{u, trub}/\delta \theta$, where $f$ is a fudge factor that is usually considered to be 0.5 \citep{ostriker_2001, heitsch_2001, padoan_2001}. With the DCF method we estimate that \VA\ = 3.7 \kms, which yields a period for the fast mode equal to 0.26 Myr. Thus, both the \cite{skalidis_tassis_2021} and the DCF method suggest that \tfast\ > \tco\ when \tco\ = 1 Myr. On the other hand, when \tco\ = 0.03 Myr, we find that \tco\ is always shorter than \tfast which means that in this case fast modes could have contributed to the CO enhancement. Since we do not have precise estimates on \tco, we cannot assess whether fast modes play a role in the enhancement of the CO abundance.

\subsubsection{Uncertainties on \tslow}

Observationally, it is challenging to constrain \TK\ and thus $\rm c_{s}$. \tslow\ scales with \TK\ as \tslow\ $\propto 1/\sqrt{\rm T_{K}}$, due to the square root dependence of $\rm c_{s}$ on \TK. Even if we consider a \TK\ as high as 100 K, we obtain that $\rm c_{s} < 1$ \kms, which yields that \tco\ < \tslow, hence slow modes are slower than the CO characteristic timescale. In order to make \tco\ > \tslow, which implies that slow modes could not enhance the CO formation, we need to increase \TK\ to a few hundred K, which is unreasonably high for our target region (Appendix~\ref{sec:co_ratio}). Therefore, we confidently estimate that \tco\ < \tslow, which means that slow modes could have enhanced the formation of CO.

\subsection{Alternative scenarios}

\cite{chen_2017_striations} proposed an alternative formation scenario for striations based on colliding flows. Colliding flows form a dense post-shock sheet-like layer. Within the post-shock layer secondary oblique shocks develop. Secondary flows form a dense, stagnated sub-layer within the primary layer. If the velocity flows within the sub-layer vary perpendicular to the post-shock magnetic field, then the sub-layer rolls around its primary axis, which tends to be parallel to the mean magnetic field orientation. When the sub-layer is viewed perpendicular to the magnetic field lines, the column density map exhibits properties similar to striations: 1) there are structures aligned with the magnetic field, and 2) the periodicity of these structures is perpendicular to the magnetic field orientation. According to the model of  \cite{chen_2017_striations}, striations are not actual density enhancements but column density effects, which are imprinted in the column density maps due to the different path lengths of the corrugated post--shock sub--layers. 

In the Polaris Flare, turbulence is compressible, since \MS\ > 1. Thus, shocks should develop within the cloud. However, 
the quasi-periodicity of CO studied here is parallel to the magnetic field which is distinct from the previously studied linear structures with axes parallel to the magnetic field but periodic spacing perpendicular to the magnetic field \citep[e.g.][]{goldsmith_2008}. \cite{chen_2017_striations} focused on the formation of structures showing quasi-periodicity perpendicular to the magnetic field, and hence we are not aware if their model can explain the quasi-periodic pattern studied here. However, the observed periodicity in the CO intensity map is unlikely to be a column density effect. The reason is that we have detected \COThirteen\ towards the $^{12}$CO intensity peaks, and inferred that the density should be maximum there compared to the rest of the region (Appendix~\ref{sec:co_ratio}). As a result, the observed CO intensity enhancements correspond to actual density increases 
and not to density corrugations as proposed by \cite{chen_2017_striations}.

An alternative scenario could be that the CO intensity variations correspond to temperature fluctuations. Consider a polytropic equation of state (EoS) for the gas, $T \propto \rho^{\Gamma}$ \citep{federrath_2015}. When $\Gamma = 1$, the gas is isothermal; when $\Gamma < 1$, small-scale fragmentation takes place, while when $\Gamma > 1$ density fluctuations are smoothed out. According to the EoS, temperature and density are two sides of the same coin. In the Polaris Flare, along with the CO intensity quasi--periodic fluctuations, we observe periodic fluctuations of the dust intensity in the three SPIRE bands (Fig.~\ref{fig:co_moment_maps}). If the dust intensity fluctuations were due to temperature fluctuations, then due to the polytropic EoS, the temperature changes 
would induce density fluctuations as well. However, the opposite is not always true, because the medium can be isothermal ($\Gamma = 1$). Overall, periodic temperature fluctuations will always be accompanied by density fluctuations, and hence the observed quasi--periodicity in the CO intensity map cannot be solely due to temperature fluctuations. 

\section{Conclusions}
\label{sec:conclusions}

We have targeted the Polaris Flare region, which is a diffuse molecular cloud with prominent striations. We used the stellar polarization data from \cite{panopoulou_2015} in order to study the magnetic field properties of the striations in the Polaris Flare. We conducted a CO (J=2-1) survey towards the striations region in order to probe the molecular gas properties and identified two regions (Region I and II) with distinct kinematics properties. The CO intensity 
enhancements are observed where dust intensity is maximum. However, dust and CO intensity structures are not perfectly correlated.   

We used the dispersion of polariation angles in order to estimate the POS magnetic field strength of the two regions with the method of \cite{skalidis_tassis_2021}. We developed a Bayesian approach (Sect.~\ref{subsec:intrnsic_spread}) in order to retrieve the intrinsic dispersion of the polarization angle distributions. For Region II, we estimate that turbulence is sub-Alfvénic with Alfvénic Mach number \MA\ $\approx 0.94 \pm 0.04$, Alfvénic speed \VA\ $\approx 2.0 \pm 0.1$ \kms, and POS magnetic field strength \Bpos\ = 38 - 76 \muG. The sonic Mach number of the region is estimated to be \MS\ = 3.9 - 6.5. All the estimated parameters are summarized in Table~\ref{table:B_properties}. 

In the CO integrated intensity map we observe a quasi--periodic structure which is almost perpendicular to the mean magnetic field orientation. The periodicity axis is $17 \degr$ off from the POS magnetic field orientation and its wavelength is 1 pc. This is in striking contrast to previous observations of striations where CO quasi-periodic structures are aligned with the magnetic field, and the periodicity axis is perpendicular to it \citep[e.g.][]{goldsmith_2008}. This quasi--periodic pattern is most likely the imprint of a travelling, and compressible MHD wave. The contrast in the CO intensity maps is larger than the contrast in the \NHmol\ map, which indicates that the CO abundance is enhanced. We compare the characteristic timescales of CO formation with the estimated dynamical timescales of MHD waves. Alfv\'en waves can enhance the CO abundance via diffusion processes \citep{federman_1996}, but this is unlikely for the Polaris Flare where the observed quasi-periodic CO intensity is correlated with dust intensity (Fig.~\ref{fig:co_moment_maps}). We find that a slow mode with velocity 0.3 - 0.45 \kms, and period 2.1 - 3.4 Myr could have enhanced the CO abundance. However, given the \tco\ uncertainties, the fast magnetosonic modes could also have played some role in the CO enhancement of the target region.

\begin{acknowledgements}
We thank the referee, Daniel Seifried, for very constructive comments which improved the manuscript. We thank J. Bieging for advice on the CO data reduction. We gratefully acknowledge the large observing time commitment provided by the Arizona Radio Observatory, and the help of the staff in conducting observations. This work was supported by NSF grant AST-2109127. AT acknowledges support by the Ambizione grant no. PZ00P2\_202199 of the Swiss National Science Foundation (SNSF). This project has received funding from the European Research Council (ERC) under the European Unions Horizon 2020 research and innovation programme under grant agreement No. 771282. KT acknowledges support from the Foundation of Research and Technology - Hellas Synergy Grants Program through project POLAR, jointly implemented by the Institute of Astrophysics and the Institute of Computer Science. GVP acknowledges support by NASA through the NASA Hubble Fellowship grant  \#HST-HF2-51444.001-A  awarded  by  the  Space Telescope Science  Institute,  which  is  operated  by  the Association of Universities for Research in Astronomy, Incorporated, under NASA contract NAS5-26555. This research was carried out in part at the Jet Propulsion Laboratory, which is operated by the California Institute of Technology under a contract with the National Aeronautics and Space Administration (80NM0018D0004). This research has made use of data from the Herschel Gould Belt survey (HGBS) project (\url{http://gouldbelt-herschel.cea.fr}). The HGBS is a Herschel Key Programme jointly carried out by SPIRE Specialist Astronomy Group 3 (SAG 3), scientists of several institutes in the PACS Consortium (CEA Saclay, INAF-IFSI Rome and INAF-Arcetri, KU Leuven, MPIA Heidelberg), and scientists of the Herschel Science Center (HSC).
\end{acknowledgements}

\bibliographystyle{aa}
\bibliography{bibliography}

\begin{appendix}
 
\section{Identifying blended stars with Convolutional Neural Networks}
\label{sec:CNN_model}

In order to eliminate blended objects in the Robopol  images we employed the CNN strategy developed by \cite{paranjpye_2020}. The CNN model consists of three convolution and max-pooling layers. For the CNN training, we used data from the archive of RoboPol \citep{blinov_robopol_archive}. 

We trained the CNN in order to identify blended spots. We flagged spots of different stars as "blends" or "normal" by visually inspecting the photometry curve of each spot. Photometry curves show the total photon flux versus the aperture size within which the photometry is performed (Fig.~9 in \citealt{panopoulou_2015}). When a star behaves "normally", the total intensity of a spot converges to a constant value at large radii, while photometry curves of blended stars do not converge at large radii. The reason is that at large radii adjacent sources contaminate the flux of the target source, and hence the photon intensity grows non-linearly. 

\begin{figure}
       \centering
       \includegraphics[width=\hsize]{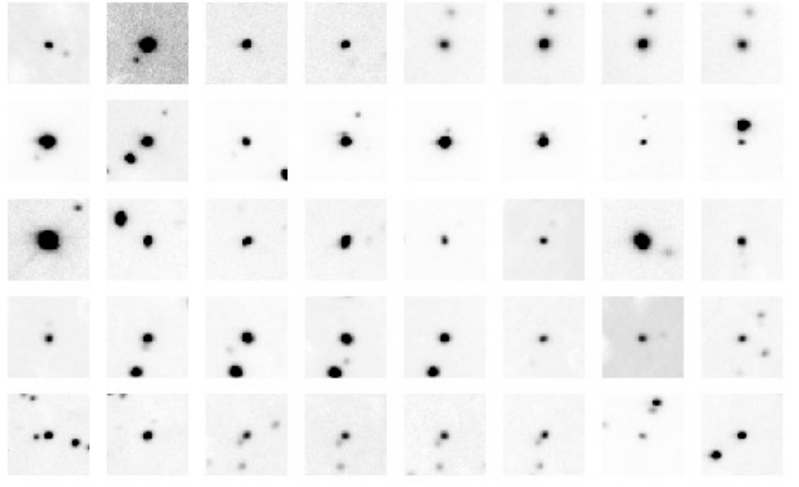}
       \caption{Images showing blended objects used for the training of the CNN. The size of each image is $64\times64$ pixel.}
       \label{fig:training_sample}
   \end{figure}

In total, our training sample consisted of 896 blended and 896 normal stars. In order to increase the size of the training sample, we flipped vertically every image. The flipped image is non-identical to the un-flipped image, but the flagged class (blend or normal) remains the same. In Fig.~\ref{fig:training_sample} we show examples of blended spots used for the training of the CNN. The classified spot is located at the center of each image.

\begin{figure}
       \centering
       \includegraphics[width=\hsize]{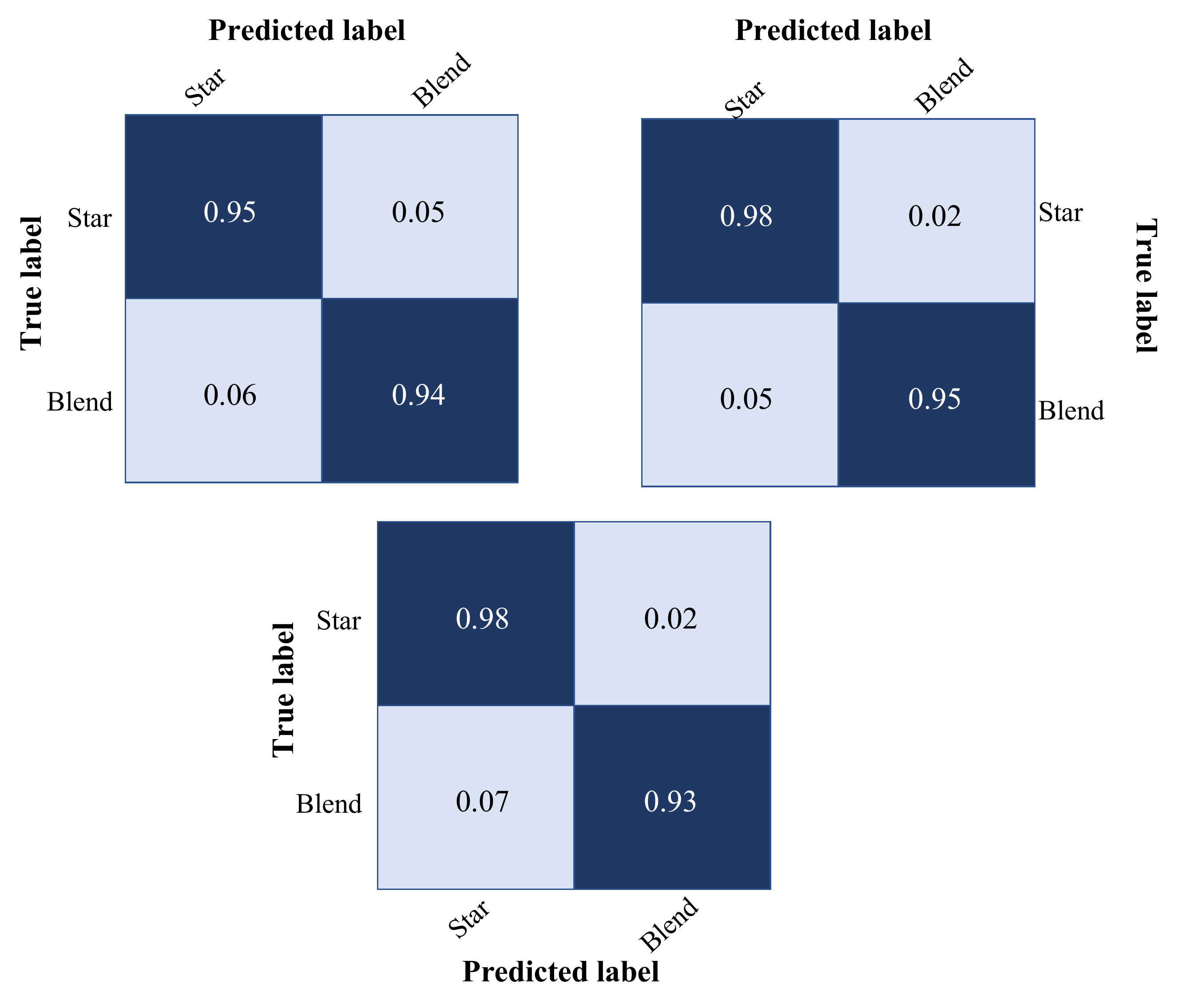}
       \caption{Normalised confusion matrices of the three neural networks trained for the identification of blended stars in polarization images. The probability of true--positives, and false--negatives is larger than 0.9 for every model. The probability of true--negatives, and false-positives is less than 0.1 for every model.}
       \label{fig:confusion_matrix}
   \end{figure}

We used 10 images, which were not included in the training sample, from both classes (normal, and blends) in order to test the accuracy of the trained CNN models. We trained the CNN model with a fixed training sample several times, and found that the accuracy of the model fluctuated statistically. In order to eliminate these statistical fluctuations we trained three different CNN models. These models were chosen due to their low false--positive, and false--negative rates. The confusion matrix of the models is shown in Fig.~\ref{fig:confusion_matrix}. We applied each model to the polarization data, and obtained three different samples of normal stars; one sample from every CNN. Overlapping measurements from each normal-star sample were considered only ones in the final sample. 

After applying the three CNN models, we could still identify some outliers in the sample. Outliers can be easily identified because their polarization angle orientations are significantly different than the local averaged polarization orientations, and also tend to have larger polarization fractions than the local averaged value \citep[e.g,][]{panopoulou_2015}. 

In order to remove outliers, we employed the relation between the maximum degree of polarization ($p_{\rm max}$) and dust extinction, \EBV, as was done in \cite{panopoulou_2015}. 
The $p_{\rm max}$ versus \EBV\ relation is empirical and indicates that at a given \EBV, ISM-induced dust polarization has an upper limit in $p$. The upper limit has been widely studied during the past years \citep{andersoon_2007, frisch_2015,panopoulou_2015,skalidis_2018, planck_coll_2020_ebv}. The most recent updated relation is $p (\%) \leq 13\times\rm E(B-V)$ \citep{panopoulou_2019_p_extinction}; measurements with S/N $\leq 3$ were debiased following \cite{vaillancourt_2006}. All the measurements which exceed the $p$ versus \EBV\ limit are likely to be intrinsically polarized sources, or blended measurements which were not identified by the CNN, hence were excluded from the analysis. We computed the extinction of each star from the 3D map of \cite{green_2018}. We extracted the distance of each star from the geometric distances from the catalogue of \cite{bailer_jones_2021}, which includes measurements from \textit{Gaia} DR3 \citep{gaia_dr3}. After we applied the $p$ - E(B-V) filtering, the remaining sample consisted of 631 measurements. Finally, we included all the normal stars from the sample of \cite{panopoulou_2015} which were excluded by our CNN models as true-positives. The final sample consists of 711 reliable polarization measurements, which is larger than the original sample of \cite{panopoulou_2015}, which consists of 609 measurements. All stars are located behind the cloud, hence their polarization signal probes the dust, and magnetic field properties of the Polaris flare.

In Fig.~\ref{fig:herschel_polaris_polarization} we show the starlight polarization data (red segments) overplotted on the \textit{Herschel} emission map. The length of each segment is proportional to the debiased $p$. A scale segment is shown in the upper left corner of the figure. 

\section{The ratio between $^{12}$CO and \COThirteen}
\label{sec:co_ratio}

The relative abundance ratio of $^{12}$CO, and $^{13}$CO can be used to constrain the cloud density and temperature \citep{pety_2011}. The $^{13}$CO (J=2-1) emission is largely undetected towards the striations region. However, the few detections that we have allow us to constrain the local gas conditions.

We focus on lines of sight where we detected both the $^{12}$CO (J=2-1), and the $^{13}$CO (J=2-1) emission lines. We fit the two CO emission lines simultaneously with RADEX \citep{VanDerTak_radex}. RADEX is a publicly available code which solves the non-local-thermodynamic-equilibrium radiative transfer equations. RADEX takes four free parameters as input: 1) \TK, 2) total CO column density (\NCO), 3) \nHmol, and 4) CO linewidth. Only the last is directly observable.

For the estimation of \NCO\ we employ the correlation between \NHmol\ and \NCO, defined as \XCO=\NCO/\NHmol. At high densities, all the available carbon is locked in CO molecules and \XCO~$=1.6\times 10^{-4}$. But, such high \XCO\ corresponds to column densities larger than $\log{\rm N_{H}} > 21$ \ColDens\ \citep{gong_2018}. At column densities $20< \log{\rm N_{H}} < 21$, where the \HI~$\rightarrow$~\Hmolecular\ transition takes place, \XCO\ is nonlinear \citep{pineda_2010}. The variability of \XCO\ can be attributed to the different formation timescales of CO, and \Hmolecular\ \citep{seifried_2017}; CO timescales are shorter than \Hmolecular.

\begin{figure}
      \centering
      \includegraphics[width=\hsize]{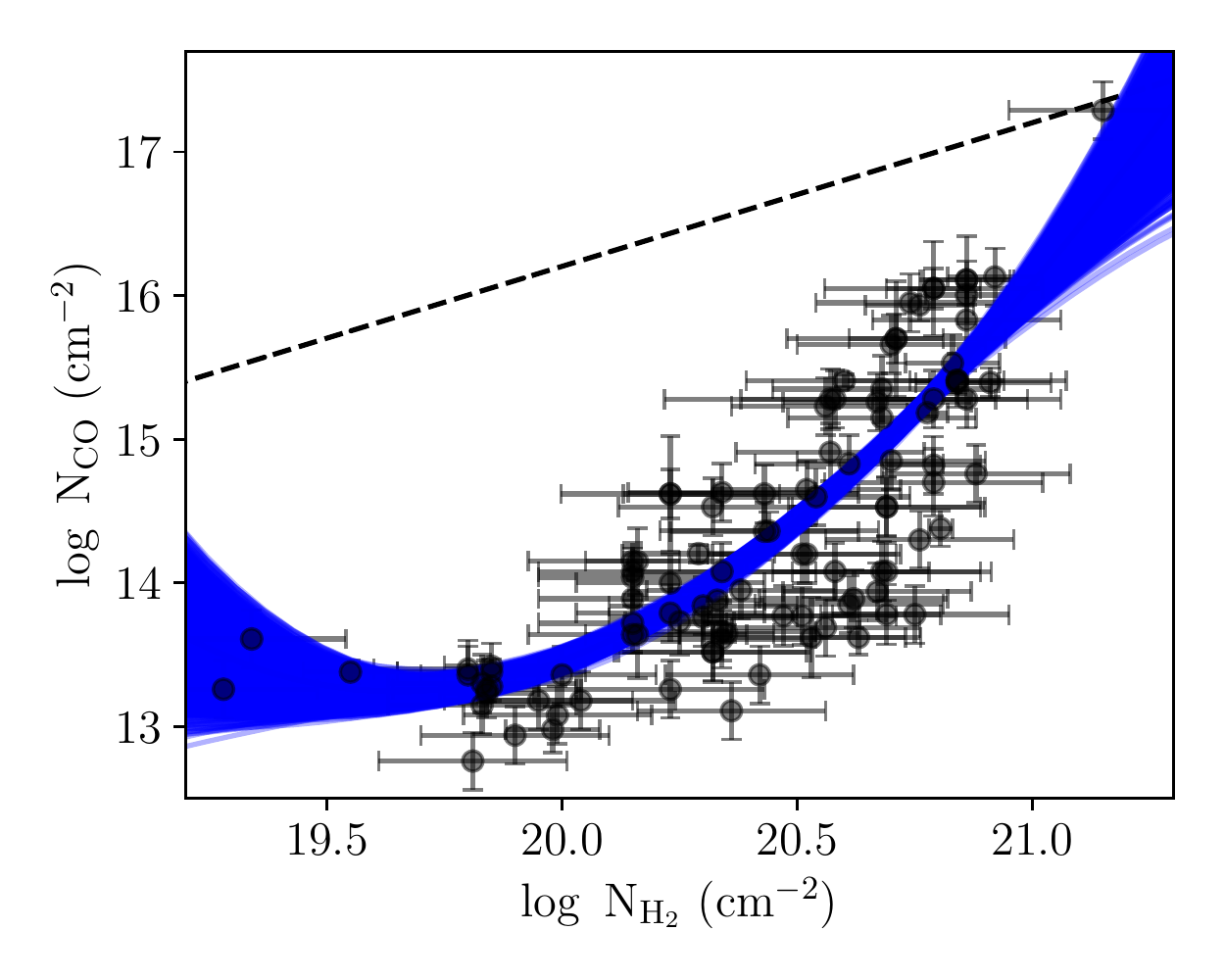}
      \caption{CO versus \Hmolecular\ column density. Black points correspond to archival data.     The blue solid lines are 7000 realizations extracted from the posterior distributions of the Bayesian fitting. The black dotted line corresponds to $10^{-4}$~\NHmol, which corresponds to clouds where essentially all carbon is in the form of CO.}
      \label{fig:NCO_NH2}
  \end{figure}

Direct constraints on \XCO\ can be obtained only with UV-absorption data \citep{burgh_2007}.  We collected archival \NCO, and \NHmol\ data \citep{burgh_2007, sheffer_2008, Gudennavar_2012}, and fitted a third degree polynomial using the Bayesian method described in Sect.~\ref{subsec:intrnsic_spread}, assuming uniform priors (Fig.~\ref{fig:NCO_NH2}). We added an extra term in the likelihood (Eq.~\ref{eq:log_likelihood}) which penalizes the observational uncertainties of \NHmol. There are no observations for $\log$ \NHmol $\geq 21.1$, but this is not a major problem because in the striations of the Polaris Flare $\log$ \NHmol $< 21$. The \XCO\ fitted relation is:
\begin{equation}
    \label{eq:nco_nh2}
    \rm \log {N_{CO}} = -0.18\left ( \log{N_{H_{2}}} \right )^{3} + 12.43\left( \log{N_{H_{2}}} \right)^{2} - 279 \log{N_{H_{2}}} + 2078.
\end{equation}
Using the \NHmol\ map of the Polaris Flare \citep{andre_2010}, we find for Region I that the median $\log~$\NHmol\ = 20.83 \ColDens, which yields, due to Eq.~(\ref{eq:nco_nh2}), that $\log~$\NCO\ $\approx$ 15.44 \ColDens. The scatter of the points in Fig.~\ref{fig:NCO_NH2} is significant, and hence we consider this value as an order of magnitude estimate.

The scatter of \NCO\ in Fig.~\ref{fig:NCO_NH2} reflects intrinsic variations of the \XCO\ relation, and is not due to observational uncertainties. The correlation between \NCO\ and \NHmol\ could be mainly affected by three physical quantities \citep{bisbas_2021}: 1) cosmic ray ionization, 2) FUV intensity, and 2) metallicity. High cosmic ray ionization would increase the \CII\ abundance \citep{goldsmith_2018}, which means that not all of the gas-phase carbon would be in the form of CO molecules. Thus, for a given column density we may end up having \CII\ associated with \Hmolecular, but not with CO; this gas component is referred to as CO-dark \Hmolecular\ \citep{van_dischoek_1988, sternberg_1995, Grenier_2005, langer_2010, paradis_2012, pineda_2013}. Large FUV intensities photodissociate CO, and lead to lower CO abundances. Finally, metallicity is linearly correlated with visual extinction, hence with \NHmol. Shielding is proportional to metallicity. Thus, CO is destroyed more efficiently in lower metallicity environments \citep{bialy_sternberg_2016}. Metallicity and FUV intensity variations could affect the column density threshold where the  \HI~$\rightarrow$~\Hmolecular\ transition takes place \citep{bialy_sternberg_2016, gong_ostriker_2020}. The points in Fig.~\ref{fig:NCO_NH2} correspond to sparsely distributed lines of sight across our Galaxy, hence each point is sensitive to the local ISM conditions. 

For the RADEX input, we assume some typical values for \TK\ and \nHmol: $\rm T_{K} = 20, 50, 100$ K, and \nHmol\ = 400, 700, 1000 \VolDens. The properties of the Polaris Flare should be represented by some pair of \nHmol, \TK; in total there are nine pairs. For each pair of \nHmol, \TK, we determine the column density of the emitted species ($^{12}$CO, and $^{13}$CO) that reproduces the observed values, and hence we obtain a pair of $^{12}$CO, and $^{13}$CO column densities (N[$^{12}$CO], N[$^{13}$CO]). Then, we calculate the ratio N[$^{12}$CO]/N[$^{13}$CO]. For every considered pair of (\nHmol, \TK), we find that N[$^{12}$CO]/N[$^{13}$CO] $\leq 10$. This ratio is significantly  
lower than the observed $^{12}$C/$^{13}$C ratio, which in our Galaxy ranges from 30 up to 70 \citep{langer_penzias_1990}. 

The reduced relative abundance of the CO isotopes {\b stronly suggests} that isotopic fractionation of CO is taking place \citep{watson_1976}. The isotopic fractionation occurs due to the exothermal ion--neutral reaction
\begin{equation}
\label{chemical_reaction:isotopic_fractionation}
   \rm ^{13}C^{+} + ^{12}CO \rightarrow ^{12}C^{+} + ^{13}CO + \Delta E,
\end{equation}
where $\Delta E/k_{B} = 35$ K. The column density range where isotopic fractionation is efficient is log \NCO (\ColDens)\ = 15 -- 16 \citep{szucs_2014}, which is consistent with our order of magnitude estimate of \NCO, based on Eq.~(\ref{eq:nco_nh2}). The above chemical reaction enhances the abundance of $^{13}$CO relative to $^{12}$CO. The carbon monoxide isotopic ratio depends on \TK\ as \citep{watson_1976}:
\begin{equation}
    \label{eq:co_isotopic_ratio_fractionation}
    \rm \frac{^{12}CO}{^{13}CO} = \frac{^{12}C}{^{13}C} \times  e^{-35/T_{K}}.
\end{equation}
It is evident that the isotopic fractionation is effective at low temperatures. When \TK\ $> 100$ K, the CO isotopic ratio converges to the $\rm ^{13}C/^{12}C$ ratio, which is close to 60 in our Solar neighborhood \citep{langer_penzias_1990}. As \TK\ decreases (\TK\ $\leq 50$ K), then $\rm ^{13}CO/^{12}CO$ increases exponentially (Eq.~\ref{eq:co_isotopic_ratio_fractionation}), compared to the elemental carbon ratio. 

For a given \TK, we compute analytically the expected CO isotopic ratio (Eq.~\ref{eq:co_isotopic_ratio_fractionation}), and compare this value with the RADEX-output N[$^{12}$CO]/N[$^{13}$CO] ratio. When the analytical and RADEX results agree, we consider that the input \TK\ represents the actual cloud temperature. 

The isotopic ratio of $^{12}$C/$^{13}$C depends linearly on the Galactocentric distance (D$_{\rm GC}$, \citealt{milam_2005}). For the Polaris Flare, we estimate that D$_{\rm GC} = $ 8.3 kpc, assuming for the Sun that D$_{\rm GC} = $ 8.1 kpc \citep{bobylev_2021}. Based on Eq.~(3) of \cite{milam_2005}, we estimate that $^{12}$C/$^{13}$C = 63. Without isotopic fractionation, or equally at high temperatures (\TK\ = 100 K), $^{12}$CO/$^{13}$CO should be close to 63. The analytic CO isotopic ratio (Eq.~\ref{eq:co_isotopic_ratio_fractionation}) is consistent with the ratio obtained from the RADEX-output column densities only when \TK\ $\approx 20$ K. The \nHmol\ = 400 \VolDens\ case seems to be unlikely, because when we fit the observed emission lines with this density, the data are reproduced only when N($^{12}$CO) = 3 $\times 10^{16}$ \ColDens. This value is higher than the \NCO\ values with log \NHmol(\ColDens)\ < 21  (Fig.~\ref{fig:NCO_NH2}). If we consider a maximum log \NCO(\ColDens)\ = 16  (Fig.~\ref{fig:NCO_NH2}), then we find that the minimum density that can fit the data is 600 \VolDens. Overall, regions with detectable \COThirteen\ should have \TK\ $\approx$ 20 K, and \nHmol\ $\geq$ 600 \VolDens.

\begin{figure}
      \centering
      \includegraphics[width=\hsize]{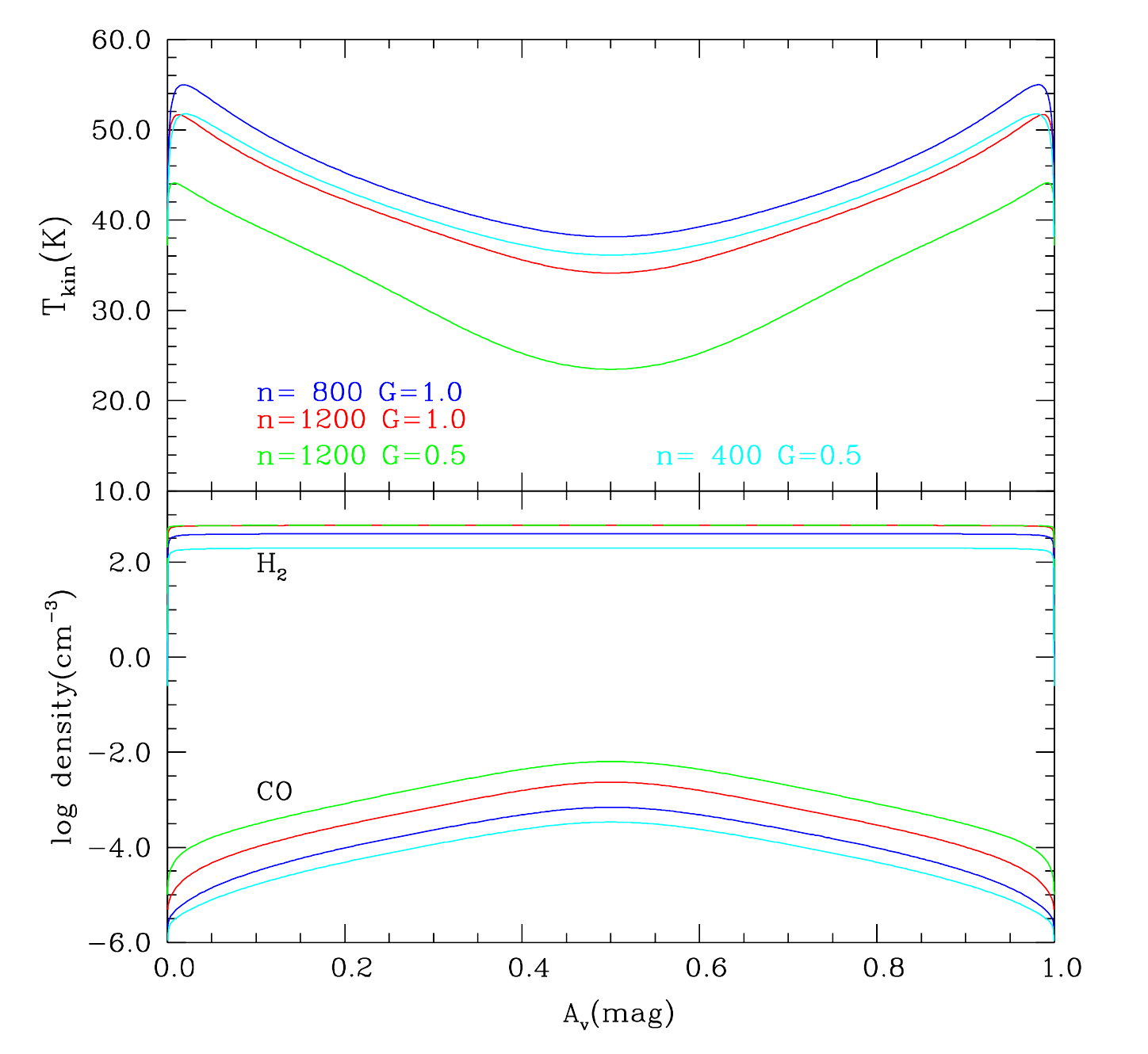}
      \caption{Results of the PDR modelling for $\rm A_{v}=1$. \textbf{Upper panel:} Kinetic temperature versus $\rm A_{v}$ for different gas densities and UV intensities ($\rm G$). \textbf{Lower panel:} Densities of \Hmolecular\ and CO versus $\rm A_{v}$.}
      \label{fig:pdr_code}
  \end{figure}

For a deeper understanding of the cloud properties, we modeled the conditions towards the regions where \COThirteen\ was detected. We employed the Meudon PDR code \citep{le_petit_2006}. We do not aim to model the average properties of the striation region, but only the specific lines of sight. We used a constant visual extinction $A_{v} = 1$, and explore the kinetic temperature as a function of density and of the UV background intensity ($\rm G$) relative to the standard Habing interstellar radiation field intensity (ISRF). The cloud is modeled as a two--sided slab with equal UV field incident on both sides.  In the upper panel of Fig.~\ref{fig:pdr_code} we show the results of the modelling. \TK\ is strongly affected by $\rm G$, and weakly by \nHmol; \nHmol\ is important only for the relative CO abundance (lower panel in Fig.~\ref{fig:pdr_code}). So far we have inferred that \TK\ should be close to 20 K. From the PDR modelling, we find that \TK\ = 20 K is achieved only for \nHmol\ = 1200 \VolDens, and $\rm G = 0.5$ 
This is an upper limit for the density of the striations region. If we assume that the mean UV intensity in the striations region is $\rm G \leq 1$, then the average kinetic temperature should be \TK\ $\lesssim 50$ K, which is consistent with the \TK\ that we assumed in Sect.~\ref{subsec:dens_estimation}.

\section{CO formation timescale}
\label{app:co_formation_timescale}

  \begin{figure}
      \centering
      \includegraphics[width=\hsize]{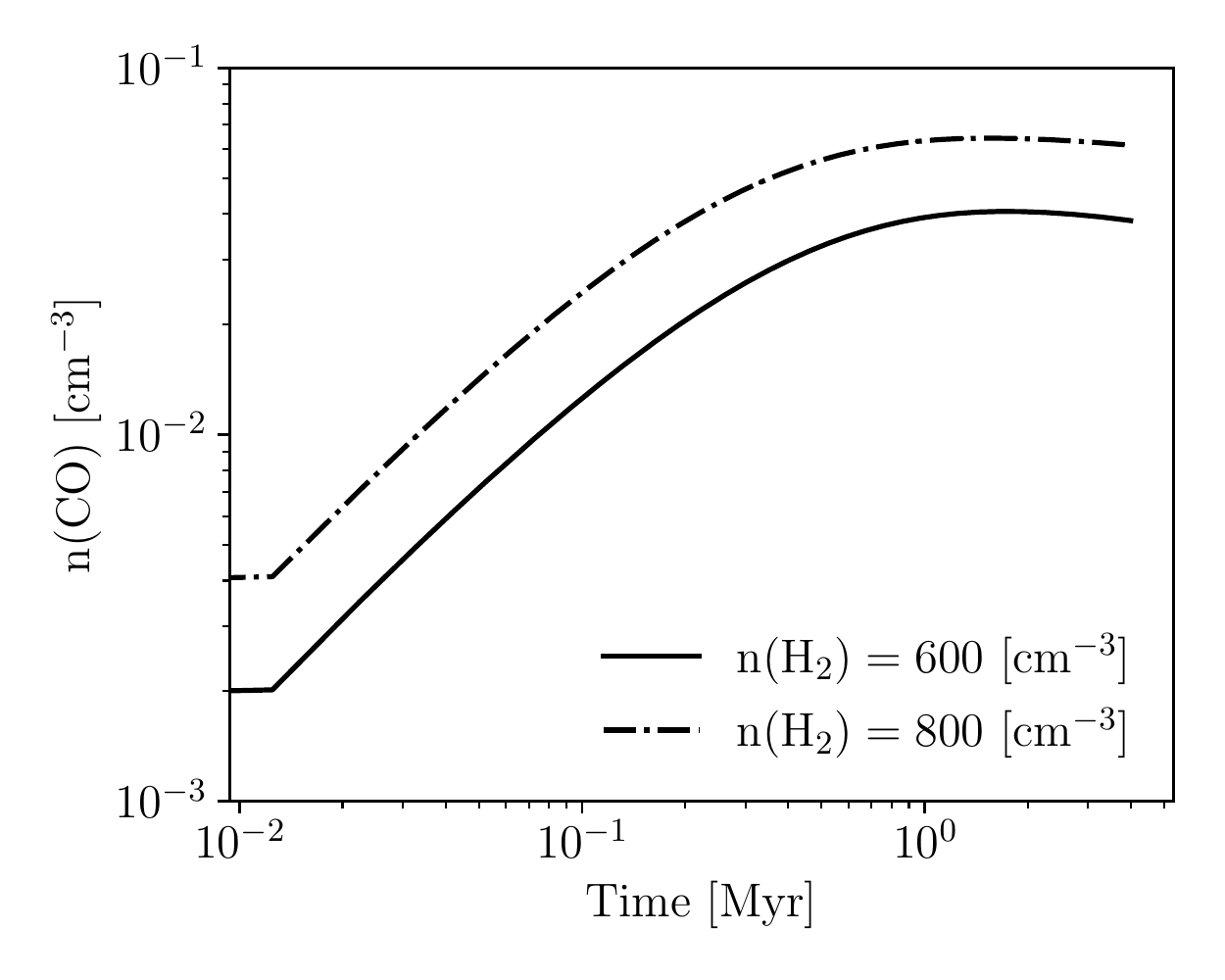}
      \caption{CO density time evolution for a simulated cloud with $\rm A_{v}=1$, and T=35 K.}
      \label{fig:co_timescale}
  \end{figure}
  
We explore the time evolution of $\rm{CO}$ under dynamical steady state. We use a modified version of the FLASH astrophysical code \citep{2000ApJS..131..273F, 2008ASPC..385..145D}, which includes a chemical network consisting of 115 chemical species and 1600 chemical reactions. Our chemical network was first used in \cite{2022MNRAS.510.4420T}, while the details of the chemical processes and the initial elemental abundances have been presented in \cite{2016MNRAS.458..789T}. We consider a cylindrical cloud geometry and explore two possibilities regarding the physical conditions of the Polaris cloud, based on our estimated densities. In the first case, the $\rm{H_2}$ number density was set to \nHmol\ = 800 \VolDens, and in the second case we assume $n_{\rm{H_2}} = 600$ \VolDens. In both cases, the temperature and visual extinction were set equal to T=35 K and $\rm A_{v}$ = 1.

In Fig.~\ref{fig:co_timescale} we show the evolution of the $\rm{CO}$ number density as a function of time. When $n_{\rm{H_2}} = 800 ~\rm{cm^{-3}}$, the $\rm{CO}$ abundance is consistently larger than that for n$_{\rm{H_2}}$ = 600 \VolDens\ but the rate of increase in $n({\rm CO})$ is essentially the same. The timescale required to reach the first maximum abundance of $n({\rm CO})$ is \tco~$\approx$ 1Myr for any n(\Hmolecular). If we induce a detectability cutoff at $\rm n_{CO}=0.01$ \VolDens, then for n(\Hmolecular) = 600 \VolDens\ we obtain that \tco\ $\approx 1$ Myr, while for n(\Hmolecular) = 800 \VolDens, \tco\ $\approx$ 0.03 Myr. In order to be conservative, we consider a large range of CO timescales, \tco\ = 0.03 - 1.0 Myr. 
\end{appendix}

\end{document}